\documentclass[11pt]{article}

\usepackage{cancel}
\usepackage{multirow}
\usepackage{cite}
\usepackage{slashed}
\usepackage{amsmath}
\usepackage{amssymb}
\usepackage{graphicx}
\usepackage[latin1]{inputenc}
\usepackage{mathrsfs}
\usepackage[dvips]{color}
\usepackage[footnotesize]{caption}

\newcommand{\sw}{s^2_W}
\newcommand{\eps}{\varepsilon}
\newcommand{\bea}{\begin{eqnarray}}
\newcommand{\eea}{\end{eqnarray}}
\newcommand{\be}{\begin{equation}}
\newcommand{\ee}{\end{equation}}
\newcommand{\ba}{\begin{array}}
\newcommand{\ea}{\end{array}}

\def\gsim{\mathrel{\rlap{\lower4pt\hbox{\hskip1pt$\sim$}}
    \raise1pt\hbox{$>$}}}

\begin{document}

\begin{titlepage}
\
\vspace*{-15mm}
\begin{flushright}
MPP-2014-313\\

\end{flushright}
\vspace*{0.7cm}

\begin{center}
{
\bf\Large
Non-unitarity of the leptonic mixing matrix: \\[1mm] Present bounds  
and future sensitivities
}
\\[8mm]
Stefan~Antusch$^{\star\dagger}$
\footnote{E-mail: \texttt{stefan.antusch@unibas.ch}}
and 
Oliver~Fischer$^{\star}$
\footnote{E-mail: \texttt{oliver.fischer@unibas.ch}},
\\[1mm]
\end{center}
\vspace*{0.50cm}
\centerline{$^{\star}$ \it
 Department of Physics, University of Basel,}
\centerline{\it
Klingelbergstr.~82, CH-4056 Basel, Switzerland}
\vspace*{0.2cm}
\centerline{$^{\dagger}$ \it
Max-Planck-Institut f\"ur Physik (Werner-Heisenberg-Institut),}
\centerline{\it
F\"ohringer Ring 6, D-80805 M\"unchen, Germany}
\vspace*{1.20cm}
\begin{abstract}
\noindent
The non-unitarity of the effective leptonic mixing matrix at low energies is a generic signal of extensions of the Standard Model (SM) with extra fermionic singlet particles, i.e.\ ``sterile'' or ``right-handed'' neutrinos, to account for the observed neutrino masses. 
The low energy effects of such extensions can be described in a model-independent way by the Minimal Unitarity Violation (MUV) scheme, an effective field theory extension of the SM. We perform a global fit of the MUV scheme parameters to the present experimental data, which yields the up-to-date constraints on leptonic non-unitarity. Furthermore, we investigate the sensitivities and discovery prospects of future experiments. In particular, FCC-ee/TLEP would be a powerful probe of flavour-conserving non-unitarity for singlet masses up to $\sim$ 60 TeV. Regarding flavour-violating non-unitarity, future experiments on muon-to-electron conversion in nuclei could even probe extensions with singlet masses up to $\sim$ 0.3 PeV.
\end{abstract}

\end{titlepage}

\setcounter{footnote}{0}

\section{Introduction}

Neutrino masses provide a clear indication that the Standard Model (SM) of elementary particles has to be extended. The question which of the various possible extensions to generate neutrino masses is the one realized in nature, is one of the big puzzles in particle physics. These extensions all have in common that they predict new interactions of the neutrinos as well as new particles beyond the ones contained in the known SM representations.  
 
One characteristic class of SM extensions, which generates the observed neutrino masses, features (some number of) fermionic singlets, which may be named ``sterile'' or ``right-handed'' neutrinos. In generic models with such fermionic singlets, the SM neutrinos (which are part of the leptonic SU(2)$_\mathrm{L}$-doublets) mix with the fermionic singlets. The full leptonic mixing matrix is thus enlarged, with details depending on the explicit model under consideration.
However, in the case that the extra singlets are heavier than the energies of a given experiment, only the light states propagate, so that an effective leptonic $3 \times 3$ mixing matrix emerges. The latter is given by a submatrix of the full unitary leptonic mixing matrix and thus in general {\em not} unitary (see, e.g., \cite{Langacker:1988ur}). The new physics effects within this class of SM extensions are well captured by an effective theory extension of the SM referred to as the Minimal Unitarity Violation (MUV) scheme \cite{Antusch:2006vwa}. 

Constraints on non-unitarity within MUV have been discussed and calculated, e.g., in \cite{Antusch:2006vwa,Antusch:2008tz}. The connection to so-called non-standard neutrino interactions (NSIs) has been discussed in \cite{FernandezMartinez:2007ms,Antusch:2008tz} and an analysis of the potential to test MUV at future neutrino oscillation facilities was performed in \cite{FernandezMartinez:2007ms,Antusch:2009pm}.  
Furthermore it has been pointed out in \cite{Antusch:2009gn} that MUV extensions of the SM with significant non-unitarity can make the leptogenesis mechanism for generating the baryon asymmetry of the universe more efficient. Recently, updated constraints from electro-weak precision observables (EWPO) have been used to calculate non-unitarity constraints in Refs.~\cite{Akhmedov:2013hec,Basso:2013jka} and it has been emphasized in \cite{Basso:2013jka} that already the present data, when a certain selection of lepton sector observables is considered, shows a significant hint for non-zero non-unitarity parameters. 

In the present paper we perform a global fit to the leptonic non-unitarity parameters in the MUV scheme, including all relevant presently available experimental data. On the one hand, this goes beyond previous studies in terms of completeness. On the other hand, the discovery of the Higgs boson and the measurement of its mass reduced the theoretical uncertainties especially of the EWPOs significantly. It is therefore of interest to update the allowed/favoured regions for the non-unitarity parameters. In addition, we discuss the sensitivities to test non-unitarity in leptonic mixing at various proposed future experiments and calculate the discovery prospects as well as the potential to set improved bounds. 

The paper is organised as follows: In section~\ref{sec:MUV} we review the MUV scheme and the used parametrisation for non-unitarity in leptonic mixing. Section~\ref{sec:observables} contains a summary of the considered observables and the modified theory predictions in MUV and in section~\ref{sec:analysis} we present the results of our global fit to the present data. In section~\ref{sec:future} the potential of future experiments to test leptonic non-unitarity is discussed. In section~\ref{sec:summary} we summarize and conclude.

\section{Minimal Unitarity Violation}
\label{sec:MUV}
\subsection{Effective theory extension of the Standard Model}

We consider an extension of the SM by two effective operators, one of mass dimension 5 and the second of mass dimension 6, which is referred to as Minimal Unitarity Violation (MUV) \cite{Antusch:2006vwa}:
\bea\label{eq:MUVmodel}
\mathscr{L}_{\rm MUV} &=& \mathscr{L}_\mathrm{SM} + 
\delta{\cal L}^{d=5} + \delta{\cal L}^{d=6} \:.
\eea
The first of these operators, $\delta{\cal L}^{d=5}$,  generates the neutrino masses after electroweak symmetry breaking, while the second, $\delta{\cal L}^{d=6}$, leads to a non-unitary effective low energy leptonic mixing matrix, as will be discussed in the following. The dimension 5 (neutrino mass) operator is given by
\bea
\delta{\cal L}^{d=5} = \frac{1}{2}\, c_{\alpha \beta}^{d=5} \,\left( \overline{L^c}_{\alpha} \tilde \phi^* \right) \left( \tilde \phi^\dagger \, L_{ \beta} \right) + \mathrm{H.c.} \;,
\label{eq:dim5op}
\eea
with $\phi$ denoting the SM Higgs field, which breaks the electroweak (EW) symmetry spontaneously after acquiring its vacuum expectation value (vev) $v_\mathrm{EW}=246.22$ GeV, using the notation 
$\tilde \phi = i \tau_2 \phi^*$, and with $L^\alpha$ (with $\alpha = 1,2,3$) being the lepton doublets. 
The second effective operator
\bea
\delta{\cal L}^{d=6} = c^{d=6}_{\alpha \beta} \, \left( \overline{L}_{\alpha} \tilde \phi \right) i \cancel{\partial} \left(\tilde \phi^\dagger L_{ \beta} \right)
\label{eq:dim6op}
\eea
generates, after EW symmetry breaking, a contribution to the kinetic terms of the neutrinos. Canonically normalising the kinetic terms, which in general involves a transformation of the neutrino fields by a non-unitary matrix, leads to a non-unitary leptonic mixing matrix (see, e.g., Refs.~\cite{DeGouvea:2001mz,Broncano:2002rw,Antusch:2006vwa}).

\subsection{The Standard Model plus fermionic singlets}\label{sec:SMplusSinglets}
The effective extension of the SM defined in eq.~(\ref{eq:MUVmodel}) indeed captures a quite  generic situation in a model-independent way, namely all models where the SM neutrino degrees of freedom mix with other neutral fermionic fields which are heavy compared to the energy scale where the considered experiments are performed. 

In particular, it provides an effective description of models with an arbitrary number of sterile (right-handed) neutrinos (when they are comparatively heavy). 
Indeed, the MUV scheme includes exactly the leading effective operators generated by the generalized seesaw extension of the SM, given by 
\bea\label{eq:The3FormsOfNuMassOp}
\mathscr{L} = \mathscr{L}_\mathrm{SM} -\frac{1}{2} \overline{\nu_\mathrm{R}^I} M^N_{IJ} \nu^{c J}_\mathrm{R} -(Y_{N})_{I\alpha}\overline{\nu_\mathrm{R}^I} \widetilde \phi^\dagger
L^\alpha+\mathrm{H.c.}\; ,
\eea
where the  $\nu_\mathrm{R}^I$ ($I=1,...,n$) are gauge singlet fermions, i.e.\ ``sterile'' or ``right-handed'' neutrinos. 
When the $\nu_\mathrm{R}^I$ are integrated out of the theory below their mass scales, the effective operators $\delta{\cal L}^{d=5} $ and $\delta{\cal L}^{d=6} $ are generated. The coefficient matrix in the definition of $\delta{\cal L}^{d=5}$, (cf.\ eq.~(\ref{eq:dim5op}),) can be connected to the parameters of the seesaw extension in eq.~(\ref{eq:The3FormsOfNuMassOp}):
\bea
c_{\alpha \beta}^{d=5} = (Y_N^T)_{\alpha I} (M_N)_{IJ}^{-1} (Y_N)_{J\beta}\:.
\eea
After EW symmetry breaking, $\delta{\cal L}^{d=5}$ generates a neutrino mass matrix for the light neutrinos as
\bea\label{Eq:d5AndMnu}
(m_\nu )_{\alpha\beta}= - \frac{v_\mathrm{EW}^2}{2} c_{\alpha\beta}^{d=5}\;,
\eea 
which is nothing else than the usual seesaw formula. 
In addition, in the basis where $M_N$ is diagonal, the coefficient matrix $c_{\alpha \beta}^{d=6}$ from eq.~(\ref{eq:dim6op}) can be expressed as (cf.\ \cite{Broncano:2003fq})
\bea
c_{\alpha \beta}^{d=6} = \sum_I (Y_N^\dagger)_{\alpha I} (M_N)_{II}^{-2} (Y_N)_{I\beta}\;.
\label{eq:d=6coeffs}
\eea
After EW symmetry breaking, and performing the canonical normalisation of the neutrino kinetic terms mentioned above, the unitary mixing matrix $U$ in the lepton sector gets modified to a non-unitary one which we will call $N$. More specifically, the kinetic term of the neutrinos (before canonical normalisation, for details see \cite{Antusch:2006vwa}) can be written as
\bea
\mathscr{L}_\mathrm{kin,\nu} =  i\, \overline{\nu}_{\alpha}  \,\cancel{\partial} \,(N N^\dagger)^{-1}_{\alpha\beta}  \,\nu_{ \beta} \:,
\eea
such that we can identify
\bea
(N N^\dagger)^{-1}_{\alpha\beta}  - {1}_{\alpha\beta} = \frac{v^2_\mathrm{EW}}{2} c_{\alpha \beta}^{d=6}
\eea
as the contribution from the dimension 6 operator, which governs the deviation from unitarity. 

\subsection{Parametrisation of non-unitary leptonic mixing}
There are various parametrisations in use: To start with, without loss of generality, we can write $N$ as product of a Hermitean matrix $H \equiv (1 + \eta)$ and a unitary matrix $U$ (see e.g.\ \cite{FernandezMartinez:2007ms}), where it is assumed that the elements of the Hermitean $\eta$ matrix are $\ll 1$,
\be
N = (1 + \eta)\,U\;. \label{param}
\ee

Alternatively, one can write the Hermitean combination $N N^\dagger$ as (cf.\ \cite{Antusch:2006vwa})
\be\label{eq:def_NNdagger_eps} 
(N N^\dagger)_{\alpha\beta} = (1_{\alpha\beta} + \eps_{\alpha\beta})  \:,   
\ee
where now the Hermitean matrix $\eps$ (with small entries) parametrises the deviation of the leptonic mixing matrix $N$ from being unitary. The non-unitarity parameters $\eps_{\alpha \beta}$ and $\eta_{\alpha \beta}$ are related to the coefficient matrix $c_{\alpha \beta}^{d=6}$ by
\bea\label{Eq:EpsAndNonU}
\eps_{\alpha \beta} = 2 \eta_{\alpha \beta} = - \frac{v^2_\mathrm{EW}}{2} c_{\alpha \beta}^{d=6} \;,
\eea
up to higher order terms in the small elements $\eps_{\alpha \beta}$ and $\eta_{\alpha \beta}$. 

Furthermore, in the recent works \cite{Akhmedov:2013hec,Basso:2013jka} only the diagonal elements $\eps_{\alpha \alpha}$ of the matrix $\eps$ have been considered. Those were called $\eps_\alpha$ (with a single index), and a different sign convention has been chosen such that 
\be
\eps_{\alpha} = - \eps_{\alpha \alpha}\:.
\ee

Throughout this work, we will present our results using the the non-unitarity parameters $\eps_{\alpha \beta}$, however it is straightforward to translate them into any desired convention.

\paragraph{A priory constraints on the non-unitarity parameters:}
From Eqs.~(\ref{eq:d=6coeffs}) and (\ref{Eq:EpsAndNonU}) one can easily see that the diagonal elements of $\eps$ are real and satisfy
\be
\eps_{\alpha \alpha} \le 0\:.
\label{prior1}
\ee
The off-diagonal $\eps_{\alpha \beta},$ (for $\alpha\neq \beta$) can in general be complex. However, the observables we will consider in this work (and which are available up to date) are only sensitive to the modulus $|\eps_{\alpha \beta}|$. Furthermore, the moduli of the off-diagonal elements are restricted by the triangle inequality \cite{Antusch:2008tz}
\be
|\eps_{\alpha \beta}| \le \sqrt{|\eps_{\alpha \alpha}| |\eps_{\beta \beta}|}   \:.
\label{prior2}
\ee
In our analysis, we will only allow for non-unitarity parameters which satisfy the above constraints. Consequently, our analysis makes use of the following six real parameters:\footnote{Future neutrino oscillation experiment could be sensitive to the phases of the off-diagonal $\eps_{\alpha \beta}$, as discussed in \cite{FernandezMartinez:2007ms,Antusch:2009pm}. }
\be
\eps_{ee} \:,\; \eps_{\mu\mu} \:,\; \eps_{\tau\tau} \:,\; |\eps_{e\mu}|  \:,\; |\eps_{e\tau}|\:,\; |\eps_{\mu\tau}|\:.
\label{MUVpars}
\ee

\section{Observable consequences of non-unitary leptonic mixing}\label{sec:observables}
As discussed above, after EW symmetry breaking the dimension 6 operator $\delta{\cal L}^{d=6}$ generates a contribution to the kinetic terms of the neutrinos, which, when canonically normalising them, induces a non-unitary leptonic mixing matrix $N$, replacing the initially unitary leptonic mixing matrix, the PMNS matrix $U$. 

Apart from modifying the coupling to the W bosons, the transformation of the neutrino fields by a non-unitary matrix also changes the couplings to the Z boson (see, e.g., Refs.~\cite{DeGouvea:2001mz,Broncano:2002rw,Antusch:2006vwa}).  
Consequently, in the MUV scheme, both, the charged and the neutral EW current, respectively, are modified to
\be
j_\mu^\pm =  \bar \ell_\alpha \gamma_\mu N_{\alpha j} \nu_j \,, \quad j_\mu^0 = \bar \nu_i \,(N^\dagger N)_{ij} \,\gamma_\mu \nu_j\:,
\label{eq:weakcurrents}
\ee
where $\ell_\alpha$ denote the charged leptons and where the indices $i,j$ indicate from now on the (light) neutrino mass eigenstates ($i,j \in\{1,2,3\}$) as in \cite{Antusch:2006vwa}. These modifications of the electroweak interactions are the imprint of leptonic non-unitarity in the MUV scheme, the observable consequences of which will be studied in the following.

\paragraph{Non-unitarity effects in linear order in $\eps_{\alpha \beta}$:}
Since we already know from existing studies (e.g.\ \cite{Antusch:2006vwa,Antusch:2008tz}) that the non-unitarity parameters $\eps_{\alpha \beta}$ are constrained to be below at least the percent level, our strategy will be to include their effects on the observables in leading linear order, unless stated otherwise. More explicitly, we will replace the couplings to the $Z$ and $W$ bosons by the MUV-modified ones  in the tree-level diagrams for a given process.\footnote{One exception will be rare charged lepton decays, where there is no tree-level SM contribution and so also the loop contribution will be considered within MUV.} The contributing diagrams at loop level will be taken as in the SM, without including the effects of non-unitarity. This is in general sufficient w.r.t.\ the present sensitivities, since including the non-unitarity also in the loop diagrams would yield doubly-suppressed effects (suppressed by the small $\eps_{\alpha \beta}$ as well as by a loop-suppression factor). 

To make this statement more explicit, let us consider the theory prediction for some observable $O$, which we may split up into a tree-level part $O^{\mathrm{tree}}$ and a loop-level part $\delta  O^{\mathrm{loop}}$. Indicating predictions derived within the MUV scheme and within the SM with the corresponding labels, we can write
\bea
O_\mathrm{MUV} &=& O_\mathrm{MUV}^\mathrm{tree} + \delta O_\mathrm{MUV}^\mathrm{loop} \nonumber \\
&=& O_\mathrm{SM}^{\mathrm{tree}}  (1 + \delta_\mathrm{MUV}^\mathrm{tree})+ \delta O_\mathrm{SM}^{\mathrm{loop}}  (1 + \delta_\mathrm{MUV}^\mathrm{loop})\:,
\eea
where we have separated the $O_\mathrm{MUV}^\mathrm{tree}$ into a SM contribution $O_\mathrm{SM}^{\mathrm{tree}}$ and a modification term $O_\mathrm{SM}^{\mathrm{tree}}  \delta_\mathrm{MUV}^\mathrm{tree}$ due to the non-unitarity in MUV, and analogously for $\delta O_\mathrm{MUV}^\mathrm{loop}$. Identifying the SM theory prediction $O_\mathrm{SM} = O_\mathrm{SM}^\mathrm{tree} + \delta O_\mathrm{SM}^\mathrm{loop}$ we can thus write
\bea
O_\mathrm{MUV} 
&=& O_\mathrm{SM}   + O_\mathrm{SM}^\mathrm{tree} \,\delta_\mathrm{MUV}^\mathrm{tree}+ \delta O_\mathrm{SM}^{\mathrm{loop}}  \, \delta_\mathrm{MUV}^\mathrm{loop} \nonumber \\
&=& O_\mathrm{SM}  + (O_\mathrm{SM} - \delta O_\mathrm{SM}^{\mathrm{loop}})\,\delta_\mathrm{MUV}^\mathrm{tree}+ \delta O_\mathrm{SM}^{\mathrm{loop}}  \, \delta_\mathrm{MUV}^\mathrm{loop} \nonumber \\
&=& O_\mathrm{SM}  (1 + \delta_\mathrm{MUV}^\mathrm{tree}) + \dots
\:,
\eea
where the dots include the doubly-suppressed terms $\delta O_\mathrm{SM}^{\mathrm{loop}}\, \delta_\mathrm{MUV}^\mathrm{tree} $ and $\delta O_\mathrm{SM}^{\mathrm{loop}} \, \delta_\mathrm{MUV}^\mathrm{loop} $. 

For the precision we are aiming at, it is thus appropriate to consider how the tree-level contribution to a given process changes in the MUV scheme, i.e.\ to calculate $\delta_\mathrm{MUV}^\mathrm{tree}$ defined by
\be
O_\mathrm{MUV}^\mathrm{tree}  \equiv O_\mathrm{SM}^{\mathrm{tree}}  (1 + \delta_\mathrm{MUV}^\mathrm{tree}) \:.
\ee
 It is even sufficient to expand $\delta_\mathrm{MUV}^\mathrm{tree}$ to leading order in $\eps_{\alpha \beta}$. We will follow this strategy below to derive constraints on the $\eps_{\alpha \beta}$ from the presently available experimental data. We note, that in order to fully exploit the sensitivities of the most precise future experiments (see section~\ref{sec:future}), the neglected terms $\delta O_\mathrm{SM}^{\mathrm{loop}}\, \delta_\mathrm{MUV}^\mathrm{tree} $ and $\delta O_\mathrm{SM}^{\mathrm{loop}} \, \delta_\mathrm{MUV}^\mathrm{loop} $ might become relevant.

\subsection{Electroweak Precision Observables}
As usual, we will express the electroweak precision observables (EWPO) in terms of the very well measured quantities \cite{Beringer:1900zz}:
\bea 
\alpha(m_z)^{-1} & = & 127.944(14) \:, \\
G_F & = & 1.1663787(6)\, \times 10^{-5} \text{GeV}^{-2} \:, \\ 
m_Z & = & 91.1875(21)\: , 
\eea
where $m_Z$ denotes the $Z$ pole mass, $\alpha$ the fine structure constant and $G_F$ the Fermi constant. While $\alpha$ and $m_Z$ are not modified in the MUV scheme, care has to be taken with $G_F$ which is measured from muon decay. Since the leptonic charged current interactions are modified according to eq.\ (\ref{eq:weakcurrents}), what is actually measured is $G_\mu$, which is related to $G_F$ (at tree-level) by
\be
 G_\mu^2 = G_F^2(NN^\dagger)_{\mu \mu}(NN^\dagger)_{ee}\;.
\label{eq:GF}
\ee
With this kept in mind we can now turn to the modifications of the tree-level relations in the presence of MUV, following the strategy described above. A summary of the experimental values for the EWPO, along with their SM and MUV prediction, can be found in Tab.~\ref{precisionobservables}.
New physics effects on the EWPO have also been discussed, e.g., in \cite{Burgess:1993vc,Loinaz:2002ep,Loinaz:2004qc}.

\subsubsection{The weak mixing angle $\theta_{W}$}
The weak mixing angle $\theta_W$ is related to $G_F$ and $\alpha$ at tree level (or in the on-shell scheme at any loop order) via 
\be
\sw c^2_W  = \frac{\alpha(m_Z) \pi}{\sqrt{2}  G_F m_Z^2}\,,
\label{eq:treelevel}
\ee
with $s_W = \sin (\theta_W),\: c_W = \cos (\theta_W)$. From this the MUV prediction for the weak mixing angle is obtained as:
\be
s_{W}^2 = \frac{1}{2}
\left[1-\sqrt{1-\frac{2\sqrt{2}\alpha \pi}{G_\mu m_Z^2} \sqrt{(NN^\dagger)_{ee}(NN^\dagger)_{\mu\mu}}}
\right]\,.
\label{eq:seff}
\ee
With eq.~(\ref{eq:def_NNdagger_eps}), the leading order expression in the non-unitarity parameters $\eps_{\alpha \beta}$ can be obtained.

\paragraph{The effective weak mixing angle $\theta_{W,\rm eff}$:} The most precise determination of the weak mixing angle comes from the measurement of the fermion specific asymmetry parameters $A^f_{FB}$ and $A^f_{LR}$, from the resonant decay rates at the $Z$ pole. The inclusion of (fermion specific) radiative corrections to the final state, yields the definition of the (fermion specific) effective weak mixing angle $(s_{W,\mathrm{eff}}^{f})^2$. Conventionally $f=\ell$ is used as reference value in precision analyses, which is also done in this work.

We will not use the data for the individual $A^f_{FB}$ and $A^f_{LR}$ in this analysis. A brief discussion can be found in the Appendix. We also performed a fit using the individual asymmetries and found that the results are not significantly affected.

\subsubsection{$Z$ decay parameters}
The  tree-level expression for the partial decay width $Z \to \bar f f$ in MUV, for $f\neq \nu$, is given by
\be
\Gamma_f =  N_c^f\frac{ G_\mu M_Z^3}{6 \sqrt{2} \pi} \frac{\left(g_{A,f}^2 + g_{V,f}^2\right)}{\sqrt{(NN^\dagger)_{\mu \mu}(NN^\dagger)_{ee}}}\,,
\label{eq:gzvis}
\ee
with the colour factor $N_c$ and where $g_{V,f},g_{A,f}$ are the usual vector and axial vector coupling constants, 
\be
g_{V,f} = T_3^f - 2 Q_f s_W^2\,, \qquad g_{A,f} = T_3^f\,,
\label{eq:leptoncoupling}
\ee
with $T_3^f$ being the third component of the isospin and $Q_f$ the electric charge of the fermion. The invisible $Z$ decay width is obtained by summing the partial decay widths $\Gamma_{\nu_i\nu_j}$ over the light neutrino mass eigenstates with indices $i,j = 1,2,3$, or, alternatively, over the light neutrino flavour eigenstates $\alpha,\beta = 1,2,3$:
\be
\Gamma_{inv}=
\frac{ G_\mu M_Z^3}{6 \sqrt{2} \pi} \frac{\tfrac{1}{2}  \sum_{i,j} \left|\left(N^\dagger N \right)_{ij}\right|^2}{\sqrt{(NN^\dagger)_{\mu \mu}(NN^\dagger)_{ee}}}
 = \frac{ G_\mu M_Z^3}{6 \sqrt{2} \pi} \frac{\tfrac{1}{2}  \sum_{\alpha,\beta} \left|\left(N^\dagger N \right)_{\alpha,\beta}\right|^2}{\sqrt{(NN^\dagger)_{\mu \mu}(NN^\dagger)_{ee}}}
\,.
\label{eq:gammainv}
\ee
From the individual decay widths, one can define the usual (pseudo) observables:
\be
R_q  =  \frac{\Gamma_q}{\Gamma_{had}}\,, \quad
R_\ell  =  \frac{\Gamma_{had}}{\Gamma_\ell}\,, \quad
\sigma_{had}^0  =  \frac{12 \pi}{M_Z^2} \frac{\Gamma_{ee}\Gamma_{had}}{\Gamma_Z^2}\,,\quad
R_{inv} = \frac{\Gamma_{inv}}{\Gamma_{\ell}}\,.
\ee
with $\Gamma_{had} = \sum_{q\neq t} \Gamma_q$ and $\Gamma_Z$ being the $Z$ boson total width. 
We note here that some past analyses seem to have mistaken the relative error of the theory prediction for $R_{inv}$ as the absolute error. Using the correct value makes a difference especially when calculating the possible constraints from future more precise EWPO measurements.

\subsubsection{W decays}
The decay width of the $W$ boson into a lepton-neutrino pair in the MUV scheme is given by
\bea
\label{Eq:Wdecay1}
\Gamma_{W\,,\alpha} = \sum_i\Gamma (W \to \ell_\alpha \nu_i ) = \frac{G_\mu M_W^3}{6 \sqrt{2} \pi}\frac{(N N^\dagger)_{\alpha\alpha}F_W(m_{\ell_\alpha})}{\sqrt{(N N^\dagger)_{ee}(N N^\dagger)_{\mu\mu}}} \,,
\eea
with
\be
F_W(m_{\ell_\alpha})=\left(1-\frac{m_{\ell_\alpha}^2}{m_W^2}\right)^2\left(1+ \frac{m_{\ell_\alpha}^2}{m_W^2}\right)\,.
\ee
We consider the two independent lepton-universality observables $R^W_{\mu e},\,R^W_{\tau \mu}$ as in Ref.~\cite{Loinaz:2004qc}, which can be constructed from the three different decay widths: 
\be
R_{\alpha \beta}^W = \sqrt{\frac{\Gamma_{W\,,\alpha} F(m_{\ell_\beta})}{\Gamma_{W\,,\beta} F(m_{\ell_\alpha})}} = \sqrt{\frac{(NN^\dagger)_{\alpha \alpha}}{(NN^\dagger)_{\beta \beta}}}\,.
\ee
These observables allow to directly constrain the ratios of the diagonal elements of $NN^\dagger$, i.e., the $\eps_{\alpha\alpha}$. We use the $W$ boson branching ratios from the PDG \cite{Beringer:1900zz}. The two independent $R^W_{\alpha \beta}$ are displayed in Tab.~\ref{tab:universality}, together with the other lepton universality observables from low energy experiments.

\subsubsection{The $W$ boson mass}
The $W$ boson mass can be inferred from
the tree-level relation $m_Z^2 c_W^2 = m_W^2$ and eq.~(\ref{eq:treelevel}):
\be
m_W^2 = \frac{\alpha \, \pi }{\sqrt{2} G_F s_W^2}\,.
\ee
It is sensitive to the MUV parameters through the measurement of the Fermi constant, eq.~(\ref{eq:GF}) and of course also through the expression for the weak mixing angle in eq.~(\ref{eq:seff}). We obtain the following MUV prediction for the $W$ mass:
\be
\frac{ [m_W^2]_{\rm MUV}}{ [m_W^2]_{\rm SM}} =  \left[\sqrt{(NN^\dagger)_{ee}(NN^\dagger)_{\mu \mu}}
\frac{[s_W^2]_{\rm SM}}{[s_W^2]_{\rm MUV}}
\right]\,,
\ee
which can be straightforwardly expressed in leading order in $\eps_{\alpha\beta}$.

\begin{table}
\begin{center}
\begin{tabular}{|l|c|c||}
\hline 
Prediction in MUV  & Prediction in the SM & Experiment \\
\hline \hline
$\left[R_\ell \right]_{\rm SM}(1-0.15 (\eps_{ee}+\eps_{\mu\mu}))$ & 20.744(11) & 20.767(25) \\
$\left[R_b \right]_{\rm SM}(1+0.03 (\eps_{ee}+\eps_{\mu\mu}))$ & 0.21577(4) & 0.21629(66) \\
$\left[R_c \right]_{\rm SM}(1-0.06 (\eps_{ee}+\eps_{\mu\mu}))$ & 0.17226(6) & 0.1721(30)\\
$\left[\sigma_{had}^0 \right]_{\rm SM}(1 - 0.25 (\eps_{ee}+\eps_{\mu\mu})-0.27 \eps_\tau)$ & 41.470(15) nb & 41.541(37) nb \\
$\left[R_{inv} \right]_{\rm SM}(1+0.75 (\eps_{ee}+\eps_{\mu\mu})+0.67 \eps_\tau )$ & 5.9723(10) & 5.942(16) \\
$[M_W]_{\rm SM}(1-0.11 (\eps_{ee}+\eps_{\mu\mu}))$ 		& 80.359(11) GeV & 80.385(15) GeV  \\
$[\Gamma_{\rm lept}]_{\rm SM}(1-0.59(\eps_{ee}+\eps_{\mu\mu}))$  	& 83.966(12) MeV & 83.984(86) MeV \\
$[(s_{W,\mathrm{eff}}^{\ell,\mathrm{lep}})^2]_{\rm SM}(1+0.71(\eps_{ee}+\eps_{\mu\mu}))$	& 0.23150(1) & 0.23113(21)  \\
$[(s_{W,\mathrm{eff}}^{\ell,\mathrm{had}})^2]_{\rm SM}(1+0.71(\eps_{ee}+\eps_{\mu\mu}))$  & 0.23150(1)	& 0.23222(27) \\
\hline
\end{tabular}
\caption{Experimental results and SM predictions for the EWPO, and the modification in the MUV scheme, to first order in the parameters $\eps_{\alpha \beta}$. The theoretical predictions and experimental values are taken from Ref.~\cite{Baak:2014ora}. The values of $(s_{W,\mathrm{eff}}^{\ell,\mathrm{lep}})^2$ and $(s_{W,\mathrm{eff}}^{\ell,\mathrm{had}})^2$ are taken from Ref.~\cite{Ferroglia:2012ir}.}
\label{precisionobservables}
\end{center}
\end{table}

\subsection{Low energy observables}

The EWPO provide powerful constraints on leptonic non-unitarity, however they are only sensitive to the combination $\eps_{ee}+\eps_{\mu \mu}$ and to $\eps_{\tau\tau}$, as can be seen directly from Tab.~\ref{precisionobservables}. It is therefore crucial to also include other types of observables in the analysis, which can provide complementary information. Various of them stem from experiments performed at comparatively low energy (in contrast to the EWPO from high energy collider experiments).

\subsubsection{Universality tests}
\label{sec:universalitytests}
Typically, ratios of the lepton-flavour specific charged current couplings $g_\alpha$ are considered as a test for leptonic universality:
\be
R_{\alpha \beta} = \frac{g_\alpha}{g_\beta}\,.
\label{def:universality}
\ee
Those observables are inferred from ratios of decay rates, such that most theory uncertainties cancel out.

Due to the modification of the charged current interaction specified in eq.~(\ref{eq:weakcurrents}), observables defined as in eq.~(\ref{def:universality}) allow to test the ratios of diagonal elements of $NN^\dagger$:
\be
R_{\alpha \beta} = \sqrt{\frac{(NN^\dagger)_{\alpha\alpha}}{(NN^\dagger)_{\beta \beta}}} \simeq 1 + {1 \over 2}\left(\eps_{\alpha \alpha} - \eps_{\beta \beta}\right)\,.
\ee
The processes with the highest precision are very useful to constrain the MUV parameters. We consider in particular lepton decays of the form $\ell_\alpha \to \ell_\beta \nu_\alpha \bar\nu_\beta$, pion and kaon decays to electrons and muons, and $\tau$ decays to kaons and pions. We do {\it not} consider here leptonic decays of $B$ and $D$ mesons (as was done e.g.\ in Ref.~\cite{Abada:2013aba} in the context of sterile neutrino models), due to the comparatively low precision of the observed branching ratios at present. For a list of the included universality observables, and recent experimental results, see Tab.~\ref{tab:universality}.

\begin{table}
\begin{minipage}{0.49\textwidth}
\begin{center}
$\begin{array}{|c|c|c|} 
\hline 
& \mbox{Process} & \mbox{Bound} \\
\hline\hline
R_{\mu e}^\ell &
\displaystyle{\frac{\Gamma (\tau \to \nu_\tau \mu \bar{\nu}_\mu )}{\Gamma (\tau \to \nu_\tau e \bar{\nu}_e )}} & 1.0018(14) \\ 
\hline 
%R_{\tau e}^\ell &
%\displaystyle{\frac{\Gamma (\tau \to \nu_\tau \mu \bar{\nu}_\mu )}{\Gamma (\mu \to \nu_\mu e \bar{\nu}_e )}} & 1.0024(21)
%\\
%\hline 
R_{\tau \mu}^\ell &
\displaystyle{\frac{\Gamma (\tau \to \nu_\tau e \bar{\nu}_e )}{\Gamma (\mu \to \nu_\mu e \bar{\nu}_e )}} 
& 1.0006(21)
\\
\hline\hline
R_{e \mu}^W &
\displaystyle{\frac{\Gamma (W \to e \bar{\nu}_e )}{\Gamma (W \to \mu \bar{\nu}_\mu )}}
& 1.0085(93)
\\
\hline
R_{\tau \mu}^W &
\displaystyle{\frac{\Gamma (W \to \tau \bar{\nu}_\tau )}{\Gamma (W \to \mu \bar{\nu}_e )}}
& 1.032(11)
\\
\hline
\end{array}$
\end{center}
\end{minipage}
\begin{minipage}{0.49\textwidth}
\begin{center}
$\begin{array}{|c|c|c|}
\hline
 & \mbox{Process} & \mbox{Bound} \\
\hline\hline
R_{\mu e}^\pi &
\displaystyle{\frac{\Gamma (\pi \to \mu \bar{\nu}_\mu )}{\Gamma (\pi \to e \bar{\nu}_e )}}
& 1.0021(16)
\\
\hline 
R_{\tau \mu}^\pi &
\displaystyle{\frac{\Gamma (\tau \to \nu_\tau \pi)}{\Gamma (\pi \to \mu \bar{\nu}_\mu )}}
& 0.9956(31)
\\
\hline 
R^K_{\tau \mu} & 
\displaystyle{\frac{\Gamma(\tau \to K \nu_\tau )}{\Gamma(K \to \mu \bar{\nu}_\mu)}}
& 0.9852(72) \\
\hline
R^K_{\tau e} &
\displaystyle{\frac{\Gamma(\tau \to K \nu_\tau )}{\Gamma(K \to e \bar{\nu}_e)}}
& 1.018(42) \\
\hline
\end{array}$
\end{center}
\end{minipage}
\caption{Experimental tests of lepton universality. {\it Left:} Low energy lepton decays and $W$ boson decays from LEP2. {\it Right:} Low energy leptonic decays of pions and kaons. The non-$W$ values are taken from Ref.~\cite{Amhis:2012bh}. The quantities $R^W_{\alpha \beta}$ are constructed following Ref.~\cite{Loinaz:2004qc}, the $W$ boson branching ratios are taken from the present PDG world averages.}
\label{tab:universality}
\end{table}

\subsubsection{Rare charged lepton decays}
\label{subsubsec:raredecays}
Charged lepton decays $\ell_\rho \to \ell_\sigma \gamma$ occur at one loop in the MUV scheme, while they are absent in the SM, because of the neutrinos being massless. The decay width for the lepton flavour violating process $\ell_\rho \to \ell_\sigma \gamma$ reads:
\be
\Gamma_{\ell_\rho\to \ell_\sigma \gamma} = \frac{ \alpha G_\mu^2 m_\rho^5}{2048 \pi}|\sum_k N_{\rho k}^{} N^\dagger_{k \sigma} F(x_k) |^2 \,,
\label{eq:raredecay}
\ee
where we have neglected terms of the order ${\cal O}((m_{\ell_\sigma}/m_{\ell_\rho})^2)$. $F(x_k)$ is a loop-function, depending on the mass ratio $x_k = m_{\nu_k}/M_W \simeq 0$, where $m_{\nu_k}$ are the mass eigenvalues of the light neutrinos. This allows the excellent approximation
\be
\sum_k N_{\rho k}^{} N^\dagger_{k \sigma} F(x_k) \simeq F(0)\, \eps_{\rho \sigma}\,,\quad \text{with} \quad F(0) = {10 \over 3}\,.
\ee
For $\mu \to e \gamma$, the decay width in eq.~(\ref{eq:raredecay}) can be converted into a branching ratio by dividing it by $\Gamma_\mu \simeq \Gamma(\mu \to e \bar\nu_e \nu_\mu)$. To leading order in the MUV parameters, we obtain the well known expression
\be
Br_{\mu e} = \frac{100 \alpha}{96 \pi} \left|\eps_{\mu e}\right|^2\,.
\label{eq:brmue}
\ee
For the processes involving $\tau$ decays, the relation between decay width $\tau \to \ell \bar\nu_\ell \nu_\tau,\, \ell=\mu,e$ and the total decay width is given by $\Gamma_\tau = \Gamma(\tau \to \ell \bar\nu_\ell\nu_\tau)/Br_{\tau\to\ell\bar\nu_\ell\nu_\tau}$. Together with the phase space factor for the muon $\approx 1- 8 m_\mu^2/m_\tau^2,$ and the values for the leptonic branching ratios $Br_{\tau\to\ell\bar\nu_\ell\nu_\tau}$ (see e.g.\ \cite{Beringer:1900zz}), the branching ratios for the rare tau decays can be expressed numerically as:
\be
Br_{\tau e} = \frac{1}{5.6}\frac{100 \alpha}{96 \pi} \left|\eps_{\tau e}\right|^2 \qquad \text{and} \qquad Br_{\tau \mu}  =   \frac{1}{5.9}\frac{100 \alpha}{96 \pi} \left|\eps_{\tau \mu}\right|^2\,.
\ee
Another rare charged lepton decay which is sensitive to the MUV parameters is $\mu \to eee$. It can be expressed by eq.~(\ref{eq:brmue}) times a factor $\alpha$ and the appropriate three-body phase space factor ($\simeq 1$). Here we can approximate
\be
Br_{\mu eee} \approx 1.8 \times 10^{-5}\, |\eps_{\mu e}|^2\,.
\ee

\begin{table}
\begin{center}
\begin{tabular}{|c|c|c|c|}
\hline
$^A_Z$Nucleus & $Z_{eff}^N$ & $F_N(-m_\mu^2)$ & $\Gamma_{N,cap}$ [$10^{-18}$ GeV] \\
\hline
$^{27}_{13}$Al	& 11.5 & 0.64 & 0.464 \\
$^{48}_{22}$Ti	& 17.6 & 0.54 & 1.70 \\
$^{80}_{38}$Sr	& 25.0 & 0.39 & 4.62  \\
$^{121}_{51}$Sb	& 29.0 & 0.32 & 6.72 \\
$^{196}_{79}$Au	& 33.5 & 0.16 & 8.60 \\
$^{207}_{82}$Pb	& 34.0 & 0.15 & 8.85 \\
\hline
\end{tabular}
\end{center}
\caption{Values for the effective number of coherent protons $Z_{eff}^N$, the nuclear form factor $F_N(q^2)$ and the muon capture rate for different elements. Table taken from Ref.~\cite{Kitano:2002mt}.}
\label{tab-mueconversion}
\end{table}

Also the conversion of muons to electrons in atomic scattering processes can be used to constrain the MUV parameters. The energy scale for this process is $q^2=-m_\mu^2$ and it is dominated by photon exchange. At low energy, the effective coupling of the photon to the $\mu e$ current is the same as in  $\mu \to e \gamma$,
while the extra photon coupling to the nucleus is proportional to its number of protons $Z_N$. Together with nuclear effects, the conversion rate for mu-e conversion can be expressed as \cite{Alonso:2012ji}:
\be
R_{\mu e}^N =  \frac{G_F^2 \alpha^5 m_\mu^5}{8 \pi^4 \Gamma_{N,cap}} Z_{N,eff}^4 Z_N |F_N(-m_\mu^2)|^2  \left|\eps_{\mu e}\right|^2\,,
\ee
with $\Gamma_{N,cap}$ being the muon nuclear capture rate, ($Z_{N,eff}$) $Z_N$ the (effective) number of protons and $F_N(q^2)$ the nuclear form factor. The index $N$ of these nuclear parameters represents the dependence on the isotope $N$. Values for the nuclear parameters for different elements are listed in Tab.~\ref{tab-mueconversion}. For Titanium we obtain the approximate formula 
\be
R_{\mu e}^{Ti} \approx 0.0063 \, Br_{\mu e}\,.
\ee
The most stringent present bounds on charged lepton flavour violating processes and the resulting constraints for the flavour non-conserving non-unitarity parameters are listed in Tab.~\ref{tab-offdiag}. 
The possible sensitivities of future experiments, including $\mu \to eee$ and muon-electron conversion in nuclei, are discussed in section 5.3.

\begin{table}
\begin{center}
\begin{tabular}{|c|c|c|c|}
\hline
Process & MUV Prediction & 90 \% C.L. bound & Constraint on $|\eps_{\alpha \beta}|$\\
\hline\hline
$\mu  \to e \gamma$ & $2.4 \times 10^{-3} |\eps_{\mu e}|^2$ & 5.7 $\times 10^{-13}$ & $\eps_{\mu e} < 1.5 \times 10^{-5}$\\
$\tau \to e \gamma$ & $4.3 \times 10^{-4} |\eps_{\tau e}|^2$ & 1.5 $\times 10^{-8}$ & $\eps_{\tau e} < 5.9 \times 10^{-3}$\\
$\tau \to \mu \gamma$ & $4.1 \times 10^{-4} |\eps_{\tau \mu}|^2$ & 1.8 $\times 10^{-8}$ & $\eps_{\tau \mu} < 6.6 \times 10^{-3}$\\
\hline
\end{tabular}
\end{center}
\caption{Present bounds on the charged lepton flavour violating processes $\ell_\alpha \to \ell_\beta \gamma$ and resulting constraints for the flavour non-conserving non-unitarity parameters. The experimental bounds on $\mu \to e \gamma$ are from the MEG collaboration~\cite{Adam:2013mnn}, the ones on $\tau$ decays are taken from Ref.~\cite{Blankenburg:2012ex}. }
\label{tab-offdiag}
\end{table}

\subsubsection{$CKM$ unitarity}

In the SM as well as in its MUV extension, the CKM matrix is unitary. Indicating the theory values of the CKM matrix elements with a superscript ``th'', the unitarity condition for the first row reads 
\be
|V_{ud}^{th}|^2 + |V_{us}^{th}|^2 + |V_{ub}^{th}|^2  = 1\,.
\label{constr1}
\ee
Experimentally, within the SM, the world averages for the elements are \cite{Beringer:1900zz}
\be
|V_{ud}^{exp}| = 0.97427(15)\,, \quad |V_{us}^{exp}| = 0.22534(65)\,, \quad |V_{ub}^{exp}| = 0.00351(15)\,, 
\ee
and for the squared sum:
\be
|V_{ud}^{exp}|^2 + |V_{us}^{exp}|^2 + |V_{ub}^{exp}|^2  = 0.9999 \pm 0.0006 \:.
\ee
Within the SM, this provides a successful test of CKM unitarity.  

However, in the MUV scheme, the processes from which the $V_{\alpha,\beta}$ are measured get modified, as we will discuss in detail below. In general terms, we obtain that in the MUV scheme:
\be\label{eq:Vij_th_exp}
|V_{ij}^{th}|^2 = |V_{ij}^{exp}|^2 (1 + f^\mathrm{process}(\eps_{\alpha\alpha}))\:,
\ee
where $f^\mathrm{process}(\eps_{\alpha\alpha})$ depends on the process in which the CKM matrix element is measured. In the SM limit, where the $\eps_{\alpha\beta}$ are zero, $f^\mathrm{process}(\eps_{\alpha\alpha})$ vanishes. Since $V_{ub}^{exp}$ is already very small, we will neglect its modification in MUV in the following and set $V_{ub}^{th}\equiv V_{ub}^{exp} := V_{ub}$. Constraint equations for the non-unitarity parameters $\eps_{\alpha\alpha}$ can now be derived by plugging the expressions for $|V_{ij}^{th}|^2$ from eq.~(\ref{eq:Vij_th_exp}), in terms of the measured quantity and the $\eps_{\alpha\alpha}$, into eq.~(\ref{constr1}). In the following, we will discuss the measurements used in our analysis.

Let us consider first the element $V_{ud}$, which is inferred from superallowed $\beta$ decays. Since the decay rate is proportional to $G_F$ and with the factor $(NN^\dagger)_{e e}$ from the electrons in the final state, we obtain using eq.~(\ref{eq:GF})
\be
|V^{th}_{ud}|^2 = |V_{ud}^{exp,\beta}|^2 (NN^\dagger)_{\mu\mu} \,.
\label{eq:Vud}
\ee

The matrix element $V_{us}$ can be measured in hyperon, kaon or tau decays. Since the sensitivity of hyperon decays is not competitive with the other two methods, we focus on kaon and tau decays in the following. Regarding the former, the two decay modes $K \to \pi\, e \, \bar\nu_e$ and $K \to \pi\, \mu \, \bar\nu_\mu$, have the highest precision. The experimentally determined values are displayed in Tab.~\ref{tab-ckm}. For the decay modes $K \to e$ and $K \to \mu$ we get, analogously to eq.~(\ref{eq:Vud}):
\bea
|V_{us}^{th}|^2 &=& |V_{us}^{exp,K \to e}|^2(NN^\dagger)_{\mu\mu}\,,\\
|V_{us}^{th}|^2 &=& |V_{us}^{exp,K \to \mu}|^2(NN^\dagger)_{ee}\,.
\eea

\begin{table}
\begin{minipage}{0.34\textwidth}
\centering
\begin{tabular}{|l|c|}
\hline
Process                & $V_{us} f_+(0)$     \\
\hline\hline
$K_L \to \pi e \nu$     & 0.2163(6)   \\
$K_L \to \pi \mu \nu$   & 0.2166(6)   \\
$K_S \to \pi e \nu$     & 0.2155(13)  \\
$K^\pm \to \pi e \nu$   & 0.2160(11)  \\
$K^\pm \to \pi \mu \nu$ & 0.2158(14)  \\
\hline
Average                & 0.2163(5)   \\ 
\hline
\end{tabular}
\end{minipage}
\begin{minipage}{0.65\textwidth}
\begin{tabular}{|c|c|c|}
\hline
Process & $f^{\rm process}(\eps)$  & $|V_{us}|$  \\
\hline
$\frac{B(\tau \to K \nu)}{B(\tau \to \pi \nu)}$ & $\eps_{\mu\mu}$ & 0.2262(13) \\
$\tau \to K \nu	$				& $\eps_{ee} + \eps_{\mu\mu}-\eps_{\tau\tau}$	& 0.2214(22) \\
$\tau \to \ell,\,\tau \to s$ 			& $0.2\eps_{ee} - 0.9 \eps_{\mu\mu} - 0.2\eps_{\tau\tau}$	& 0.2173(22) \\
\hline
\end{tabular}
\end{minipage}
\caption{{\it Left:} Values of $V_{us} f_+(0)$, determined from different kaon decay processes. The table is taken from Ref.~\cite{Antonelli:2010yf}, and the quantity $f_+(0)=0.959(5)$ from the review~\cite{Aoki:2013ldr}. {\it Right:} Values of $V_{us}$, determined from different tau decay modes from Refs.~\cite{Follana:2007uv,Amhis:2012bh}.}
\label{tab-ckm}
\end{table}

Tau decays offer three more ways to measure $|V_{us}|$. First, the ratio of tau to kaon and tau to pion decays allows to extract $|V_{us}|/|V_{ud}|$ such that we obtain
\be
|V_{us}^{th}|^2 = |V_{us}^{exp,\tau \to K,\pi}|^2 (NN^\dagger)_{\mu\mu}  \,.
\ee
The second tau mode to be considered is the decay $\tau \to \nu_\tau K^-$. For this mode, the modifications in MUV stem from $G_F$ and from $(NN^\dagger)_{\tau \tau}$ due to the charged current interaction with tau leptons in the final state. We can thus write
\be
|V_{us}^{th}|^2 = \left|V_{us}^{exp,\tau \to \nu_\tau K^-}\right|^2\, \frac{(NN^\dagger)_{ee} (NN^\dagger)_{\mu\mu}}{(NN^\dagger)_{\tau\tau}}  \,.
\ee

The third tau mode is given by the inclusive tau partial width to strange mesons. The value for $|V_{us}|$ is extracted from the tau branching ratios via
\be
|V_{us}^{th}| = \sqrt{R_s \left[\frac{R_{had}-R_s}{|V_{ud}^{th}|^2} - \delta R_{th}\right]^{-1}}\,,
\ee
with $R_x = Br(\tau \to x)/Br(\tau \to e)$, and $\delta R_{th}=0.240\pm0.032$ \cite{Amhis:2012bh} from lattice calculations. 
Furthermore, the hadronic branching ratio is usually expressed as 
$R_{had} = 1-R_e - R_\mu$, and $R_s = (2.875 \pm 0.050)\%$ \cite{Amhis:2012bh}. $|V_{us}^{th}|$ can now be related to a measured quantity by plugging in $|V_{ud}^{th}|$ from eq.~(\ref{eq:Vud}). 

As decribed above, constraint equations on the non-unitarity parameters are obtained by plugging the various combinations of expressions for $|V_{ij}^{th}|^2$  in terms of the measured quantities and the $\eps_{\alpha\alpha}$ parameters into eq.~(\ref{constr1}). A summary can be found in Tab.~\ref{tab-ckm}.

\subsubsection{NuTeV tests of weak interactions}
The NuTeV collaboration reported a measurement of the weak mixing angle that significantly deviated from the LEP measurement \cite{Zeller:2001hh}. 
The analysis was reinvestigated by Ref.~\cite{Bentz:2009yy}, where charge asymmetry violation and an asymmetry between $s$ and $\bar s$ quarks have been included, which resulted in values for the SM parameters not in tension with other precision data. The central quantities measured at NuTeV are the ratios of neutral to charged current cross sections:
\be
R_\nu = \frac{\sigma(\nu N \to \nu X)}{\sigma(\nu N \to \mu^- X)}\,, \qquad R_{\bar \nu} = \frac{\sigma(\bar\nu N \to \bar\nu X)}{\sigma(\bar\nu N \to \mu^+ X)}\,.
\ee
We use the NuTeV results from Ref.~\cite{Bentz:2009yy} for the analysis in the following. At tree level the MUV prediction is given by
\be
\frac{ [R_i ]_{\rm MUV}}{ [R_i ]_{\rm SM}} 
= (NN^\dagger)_{\mu\mu}^2
\frac{1/2 - [s_{W}]_{\rm MUV}^2+ \frac{5}{9}(1+r_i)[s_{W}]_{\rm MUV}^4}{1/2 - [s_{W}]_{\rm SM}^2 + \frac{5}{9}(1+r_i)[s_{W}]_{\rm SM}^4}
\,,\
\ee
where the first factor stems from the fact that a muon-neutrino beam was deployed, and we used the values $r_i = 0.5$ for $R_\nu$ and $r_i = 2$ for $R_{\bar\nu}$. The NuTeV measurements for $R_\nu$ and $R_{\bar\nu}$, together with their theory prediction in the SM and in the MUV scheme, are summarized in Tab.~\ref{tab:nutev}.

\begin{table}
\begin{center}
\begin{tabular}{|l|c|c|c|}
\hline
Prediction in MUV & Prediction in the SM & Experiment \\
\hline\hline
$\left[R_\nu\right]_{\rm SM}(1 - 0.3\eps_{ee} + 1.7 \eps_{\mu\mu})$  & 0.3950(3) & 0.3933(15) \\
\hline
$\left[R_{\bar\nu}\right]_{\rm SM}(1 - 0.1\eps_{ee} + 1.9 \eps_{\mu\mu})$ & 0.4066(4) & 0.4034(28) \\
\hline
\end{tabular}
\end{center}
\caption{NuTeV results on deep inelastic scattering of neutrinos and anti-neutrinos on nuclear matter. The data has been taken from Ref.~\cite{Bentz:2009yy}. The theory uncertainty stems from $s_W^2$.}
\label{tab:nutev}
\end{table}

\subsection{Low energy measurements of $s_{W}^2$}
\label{sec:LEMs}
An important alternative approach to test the consistency of the SM parameters is given by low energy measurements, far below the $Z$ boson peak. The parity-violating nature of the weak interaction allows the measurement of the weak mixing angle with precisions below the percent level. The challenge for such measurements is the extraction of ppm to ppb asymmetries in scattering experiments, see e.g.\ Refs.~\cite{Erler:2004cx,Kumar:2013yoa} for a review.

\begin{table}[h]
\begin{center}
\begin{tabular}{|l|c|c|}
\hline 
Prediction in MUV  & Prediction in the SM & Experiment  \\
\hline \hline
$\left[ Q^{55,78}_W\right]_{\rm SM}(1+0.48 (\eps_{ee} + \eps_{\mu\mu}))$ & -73.20(35) & -72.06(44) \\
$\left[ Q^p_W\right]_{\rm SM}(1-9.1(\eps_{ee} + \eps_{\mu\mu}))$ & 0.0710(7) & 0.064(12) \\
$\left[ A_{LR}^{ee}\right]_{\rm SM}(1-15.1(\eps_{ee} + \eps_{\mu\mu}))$ & 1.520(24)$\times10^{-7}$ & 1.31(17)$\times 10^{-7}$\\
\hline
\end{tabular}
\caption{Experimental results, SM predictions and modifications in MUV for low energy experiments aiming at a measurement of $s_{W}^2$. The results on $Q^p_W$ are from Ref.~\cite{Nuruzzaman:2013bwa}. For $[A_{LR}^{ee}]_{SM}$ we used $s_W^2(M_Z)=0.2315$, and its error is dominated by the uncertainty of the radiative QED correction factors.}
\label{tab:lowE-sinW}
\end{center}
\end{table}

\subsubsection{Parity non-conservation in Cesium}
At energies far below the weak scale, an effective, parity-violating Lagrangian emerges, which describes the low-energy interactions between electrons and quarks:
\be
{\cal L}_{NC}^{eff} = -\frac{\bar e \gamma^5 \gamma^\mu e}{v_\mathrm{EW}^2}\left[g_{AV}^{eu} \frac{\bar u\gamma_\mu u}{2} + g_{AV}^{ed} \frac{\bar d\gamma_\mu d}{2}\right] - \frac{\bar e \gamma^\mu e}{v_\mathrm{EW}^2}\left[g_{VA}^{eu} \frac{\bar u \gamma^5\gamma_\mu u}{2} + g_{VA}^{ed} \frac{\bar d\gamma^5\gamma_\mu d}{2}\right]\,.
\ee
The coefficients $g_{AV}^{eq}$ and $g_{VA}^{eq}$, which define the effective quark-electron couplings, are given by products of the axial- and vector-couplings of electrons and quarks, cf.\ eq.~(\ref{eq:leptoncoupling}):
\be
g_{AV}^{eu} = -{1 \over 2} + {4 \over 3} \sw, \qquad g_{AV}^{ed} = {1 \over 2} - {2\over 3} \sw\,, 
\label{eq:effcouplings1}
\ee
and
\be
g_{VA}^{eu} = -{1 \over 2} + 2 \sw, \qquad g_{AV}^{ed} = {1 \over 2} - 2 \sw\,,
\label{eq:effcouplings2}
\ee
with $s_W^2$ being the weak mixing angle in the $\overline{MS}$ scheme. Since we are neglecting terms of the order $\delta O_{\rm SM}^{\rm loop} \times \eps_{\alpha\beta} $, we can use the tree-level relation for $\sw$ from eq.~(\ref{eq:seff}). Trough $\sw$ the low energy four fermion couplings are sensitive to the MUV parameter combination $\eps_{ee} + \eps_{\mu\mu}$. 
Alternatively, the experimental results for these couplings can be used to extract the observable $s_W^2$. 

With the definitions for the effective weak couplings of the electron and quark currents in eqs.~(\ref{eq:effcouplings1}) and (\ref{eq:effcouplings2}), the ``weak charge'' of an isotope can be defined by
\be
Q_W^{Z,N} = -2\left[Z(g_{AV}^{ep}+0.00005)+N(g_{AV}^{en}+0.00006)\right]\left(1-\frac{\alpha}{2 \pi}\right)\,,
\label{eq:qw}
\ee
where the numerical constants are due to radiative corrections in the $\overline{MS}$ scheme \cite{Erler:2013xha}, $N$ and $Z$ are the number of neutrons and protons of the isotope, respectively, and the effective electron-proton, and electron-neutron couplings are given by
\be
g_{AV}^{ep} = 2 g_{AV}^{eu}+g_{AV}^{ed}\,, \qquad g_{AV}^{en} = g_{AV}^{eu}+2 g_{AV}^{ed}\,.
\ee
The experimental measurement of $Q_W^{Z,N}$ with the highest precision has been achieved with $^{133}$Cs \cite{Dzuba:2012kx}. We display the SM and MUV prediction, and the measurement of $Q_W^{55,78}$, in Tab.~\ref{tab:lowE-sinW}.

\subsubsection{Weak charge of the proton}
\label{weakchargeproton}
The weak charge of the proton, $Q_W^p$, is obtained from eq.~(\ref{eq:qw}) by setting $Z=1,\,N=0$. Due to the smallness of this quantity, it is very sensitive to the exact value of the weak mixing angle. Note that the related observable, $Q_W^n$ is not affected by a change of $s_W^2$. 

The Qweak experiment at Jefferson lab measured $Q_W^p$ via the parity-violating asymmetry in $ep$ elastic scattering at center of mass energies of $Q^2 = 0.025$ (GeV)$^2$ until May 2012. Presently, the total dataset is being analysed, while preliminary results from $\sim$ 4\% of the total data have been published in Ref.~\cite{Nuruzzaman:2013bwa}. The extraction of the weak mixing angle, extrapolated to the $Z$ mass, yields $s^2_{W} = 0.235 \pm 0.003$.

The Jefferson Lab PVDIS Collaboration recently also published the results from an electron-quark scattering experiment. The extraction of the weak mixing angle, extrapolated to the $Z$ mass, yields a value of $s^2_{W} = 0.2299 \pm 0.0043$ \cite{Wang:2014bba}. The analysis makes use of the results from the Qweak collaboration~\cite{Nuruzzaman:2013bwa} and also of the measurements of the weak charge of Cesium~\cite{Dzuba:2012kx}.

Since the second measurement is not independent and we have no information on the correlations, we display only the preliminary Qweak results on the weak charge of the proton, together with the SM and MUV prediction in Tab.~\ref{tab:lowE-sinW}.

\subsubsection{M$\o$ller scattering}
The parity-violating part of the electron-electron interaction is to leading order a purely weak neutral current process. It is described by the following effective Lagrangian \cite{Erler:2013xha}:
\be
{\cal L}_{NC}^{ee} = -  \bar e \gamma^\mu \frac{g_{VV}^{ee}-g_{AA}^{ee}\gamma^5 e \bar e \gamma^5 + 2 g_{VA}^{ee} e \bar e \gamma^5}{4\,v_\mathrm{EW}^2}\gamma_\mu e\,,
\ee
where the SM tree-level relations for the parity-conserving, effective low energy four fermion couplings are given by
\be
g_{VV}^{ee} = {1 \over 2}(1-4 s_W^2)\,, \qquad g_{AA}^{ee} = {1\over 2}\,,
\ee
while the parity-violating effective low energy four fermion couplings are
\be
g_{VA}^{ee} = {1 \over 2} - 2\,s_W^2\,.
\ee
In the MUV scheme, $v_\mathrm{EW}$ has to be expressed in terms of $G_F$ from eq.~(\ref{eq:GF}), and for the weak mixing angle we have to use the definition from 
eq.~(\ref{eq:seff}).
The SLAC-E158 experiment has measured the parity-violating coefficient $g_{VA}^{ee}$ in fixed target polarized M$\o$ller scattering \cite{Anthony:2005pm}. The left-right cross-section asymmetry $A_{LR}^{ee}$ reduces to an interference term of the electroweak with the QED amplitude. For large incident electron energies $E_e$, the asymmetry is given by
\be
A_{LR}^{ee} = \frac{2 m_e E_e}{v_\mathrm{EW}^2} \frac{g_{VA}^{ee}}{4 \pi \alpha} {\cal F}^{ee}\,,
\label{eq:Amoller}
\ee
with $m_e$ being the electron mass, and ${\cal F}^{ee}\simeq 0.84$ for the experimental setup at SLAC-E158. Then, either by extracting the weak mixing angle from the experimental measurement of $A_{LR}^{ee}$, or, alternatively, considering the theory prediction for the asymmetry, the measurement can be used to compare the predictions from SM and MUV. 

The result from SLAC-E158 corresponds to $(s_{W,\mathrm{eff}}^{\ell})^2 (0.161$ GeV) $=$ $ 0.2395(15)$ \cite{Kumar:2007zze}, which can be extrapolated to the $Z$ mass: $(s_{W,\mathrm{eff}}^{\ell})^2$($M_Z$) = $0.2329(20)$ \cite{Kumar:2013yoa}. We choose to compare the observable $A_{LR}^{ee}$, of which the experimental measurement and the predictions in the SM and MUV scheme are listed in Tab.~\ref{tab:lowE-sinW}.

\section{Analysis}\label{sec:analysis}

We perform a Markov Chain Monte Carlo (MCMC) fit of the six non-unitarity parameters given in eq.~(\ref{MUVpars}) to the total of 34 observables discussed in the previous section, taking into account the modification of the SM theory predictions to leading linear order. The used $\chi^2$ has the form
\be
\chi^2(\eps) = \sum_{i,j} \Delta_i V_{ij} \Delta_j\,, \qquad \text{with}\qquad \Delta_i = \frac{O_{i,\rm MUV}(\eps) - O_{i,exp}}{\sqrt{\delta_{i,\rm th}^2 + \delta_{i,\rm ex}^2}}\,,
\label{eq:chi2}
\ee
where $V$ is the inverse of the correlation matrix $R$, as specified in sec.~\ref{sec:correlations} of the Appendix,  $O_i$ is one of the observables considered, and $\delta_{i, \rm th}$ and $\delta_{i, \rm ex}$ are the corresponding theoretical and experimental uncertainties, respectively. The symbol $\eps$ represents the MUV parameters of eq.~(\ref{MUVpars}).

\subsection{Results: Constraints on non-unitarity}
\label{sec:constraints}

From our analysis we obtain the following highest posterior density (HPD) intervals at 68\% ($1\sigma$) Bayesian confidence level (CL):
\be
  \begin{array}{ccl} \epsilon_{ee}  & = &-0.0012 \pm 0.0006 \\ | \epsilon_{\mu\mu} | & < & 0.00023 \\ \epsilon_{\tau\tau} & = & -0.0025 \pm 0.0017 \end{array} \qquad  \begin{array}{ccl} | \epsilon_{e\mu} | & < & 0.7 \times 10^{-5} \\ | \epsilon_{e\tau} | & < & 0.00135 \\ | \epsilon_{\mu\tau} | & < & 0.00048 \,,\end{array}
\label{MUVresult}
\ee
The best fit points for the off-diagonal $\eps_{\alpha\beta}$ and for $\eps_{\mu\mu}$ are at zero.  
At 90\% CL, the constraints on $NN^\dagger$ are:
\be
\left| NN^\dagger \right| = \left( \begin{array}{ccc} 0.9979 - 0.9998 & < 10^{-5} & < 0.0021 \\
< 10^{-5} & 0.9996 - 1.0 & < 0.0008 \\
< 0.0021 & < 0.0008 & 0.9947 - 1.0 \end{array}\right)\,.
\ee

We display the contributions to the total $\chi^2$ for each observable in the SM and the MUV scheme, as well as the individual $\chi^2$-difference between the two theories, in Fig.~\ref{fig:HEPchi2}. For the MUV scheme, the best-fit point from eq.~(\ref{MUVresult}) is used. In Tab.~\ref{tab:fitsummary} the $\chi^2$ of MUV and SM are summarised, with the observables grouped into three sets.

\subsection{Discussion}
Among the six relevant parameters in the MUV scheme, only non-zero $\eps_{ee}$ and $\eps_{\tau\tau}$ are improving the fit. The best-fit value for $\eps_{\mu\mu}$ is zero, and the fit would even prefer it to be positive, which is not possible within MUV and negative $\eps_{\alpha\alpha}$ are indeed imposed as prior in our analysis, cf.\ eq.~(\ref{prior1}). 

Concerning the flavour-violating parameters, $\eps_{e\mu}$ is predominantly constrained by the experimental result on the rare $\mu \to e \gamma$ decay. The experimental constraints for the other two parameters $\eps_{e \tau}$ and $\eps_{\mu \tau}$ are comparable to those from the triangle inequality in eq.~(\ref{prior2}), which is implied for MUV generated by SM extensions with fermion singlets. We emphasize at this point that a future observation of rare tau decays beyond the level allowed by the triangle inequality, or a strong experimental indication of a positive $\eps_{\alpha\alpha}$, has the potential to rule out the MUV scheme as a whole.

\begin{figure}[!htbp]
\vspace{-0.25\textheight}
\begin{center}
\includegraphics[scale=0.5]{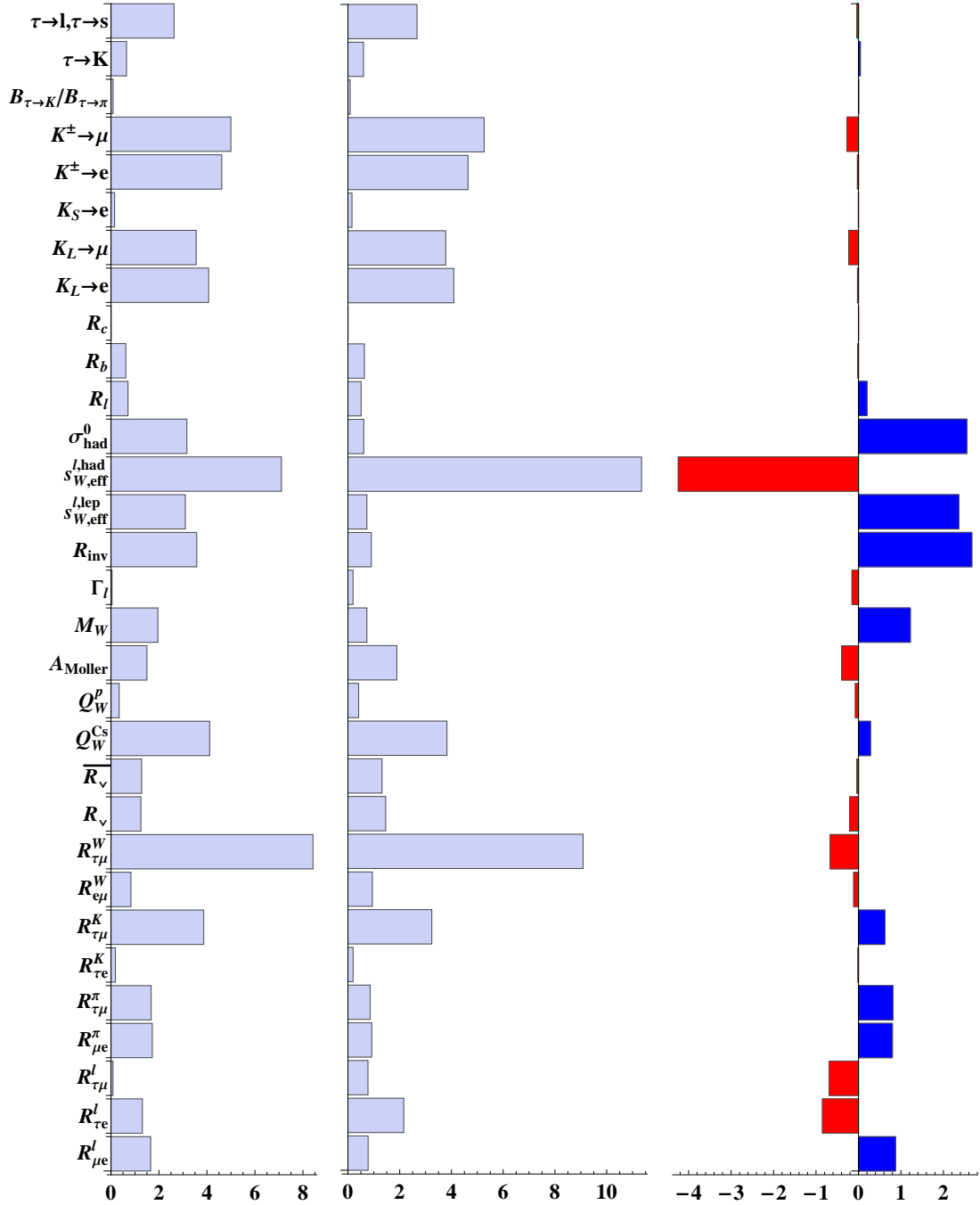}
\end{center}
\vspace{-0.22\textheight}
\caption{Individual contributions to the total $\chi^2$ from the considered observables. The left column shows the SM and the middle column the MUV scheme with best-fit parameters. The right column shows $\chi^2_i(SM)-\chi^2_i(MUV)$ for the observable $i$. The positive blue (negative red) bars indicate an improvement (worsening) of the MUV scheme best fit compared to the SM.}
\label{fig:HEPchi2}
\end{figure}

Comparing the SM with the MUV scheme at the best fit point (cf.\ table \ref{tab:fitsummary} and figure \ref{fig:HEPchi2}) we find that MUV improves the fit especially w.r.t.\ almost all EWPOs apart from $s_{W,\mathrm{eff}}^{\ell,\mathrm{had}}$, where the fit worsens from an individual $\chi^2$ of 7.1 in the SM to 11.2 in the MUV scheme. Similarly, for the universality observable $R^W_{\tau\mu}$ the individual $\chi^2$ worsens slightly from $8.4$ to $9.0$. 

The CKM observables as well as the observables from low energy experiments are not significantly changing the quality of the fit (cf.\ table \ref{tab:fitsummary}) but contribute substantially to the bounds on the $\eps_{\alpha\alpha}$. Concerning the CKM observables, we note that for an improved accuracy (especially if a strong signal for non-unitarity should be found) one should rather perform a combined fit of the CKM parameters together with the MUV parameters. This is beyond the scope of our present work. The present treatment is sufficient for deriving constraints at the given level of accuracy.

\begin{table}
\begin{center}
\begin{tabular}{|l|ccc|}
\hline
	&	SM	&	MUV	&	$\Delta \chi^2$	\\
\hline\hline
EWPO	&	20.1	&	15.3	&	4.8	\\
low energy	&	30.0	&	29.8	&	0.2	\\
CKM unitarity	&	13.7	&	14.0	&	-0.3	\\
Total	&	63.8	&	59.1	&	4.7	\\
\hline
\end{tabular}
\end{center}
\caption{Contribution to the total $\chi^2$ of the considered observables, which are grouped into three categories: The EWPO are the ones given in Tab.~\ref{precisionobservables}, the set labelled 'low energy' contains the observables from Tabs.~\ref{tab:universality}, \ref{tab-offdiag}, \ref{tab:nutev} and \ref{tab:lowE-sinW}. The CKM unitarity set is defined by Tab.~\ref{tab-ckm}. Correlations between the observables are included.}
\label{tab:fitsummary}
\end{table}

\section{Improvements from future experiments of leptonic non-unitarity tests}
\label{sec:future}
In this section we discuss how the sensitivity for tests of leptonic non-unitarity may improve with future experiments. The discussion is subdivided into four parts. The first two parts consider the improvements of the EWPO, stemming from the high energy frontier and improvements of lepton universality observables from low energy experiments and leptonic branching ratios of the $W$ boson from high energy experiments. Furthermore, we discuss the sensitivity of proposed and planned experiments for charged lepton flavour violation and also improvements from future low energy measurements of $s_W^2$. 

We will consider the discovery prospects and exclusion sensitivity of non-unitarity for the planned experimental improvements of the EWPO and the universality observables, respectively. These two sets of observables each depend on two linearly independent combinations of the MUV parameters.

\paragraph{Discovery prospects:}
We may assume that the best-fit value MUV parameters $\hat \eps$ listed in eq.~(\ref{MUVresult}) are indeed true and set 
\be
O^{exp} = O^{MUV}(\hat \eps)\,.
\label{assumption1}
\ee
To study the experimental improvement necessary for a hypothetical discovery, we consider an overall reduction of the experimental uncertainties by a factor $f_\delta = \delta^{present}/\delta^{\rm future}$, where $\delta$ denotes the experimental errors. 
The resulting function $\chi^2_{f_\delta}$ for a fixed $\eps \not= \hat \eps$ scales with $f_\delta^2$, as long as only observables are considered which are subject to the improved uncertainty.
The compatibility of the SM with the data is then given by $\chi^2_{f_\delta}(\eps=0)$. The resulting $\chi^2$-value can be used to exclude the {\it absence} of non unitarity, which can be interpreted as a discovery of non-unitarity of the effective, low energy leptonic mixing matrix. 
For a probability distribution depending on two parameters, the 5$\sigma$ discovery threshold is given by $\chi^2_{discovery} = 28.74$.

\paragraph{Exclusion sensitivity:}
In order to establish the prospects of future experiments for improving the bounds on the $\eps_{\alpha\beta}$ parameters, we assume that all non-unitarity parameters are exactly zero ($\eps \equiv 0$), and set 
\be
{\cal O}^{exp}  = {\cal O}^{SM}\,.
\label{assumption2}
\ee
As for the discovery prospects, we assume that that the experimental uncertainties are reduced by an overall factor $f_\delta$, where $\delta$ denotes the experimental errors. We display the hypothetical constraints at the 90\% Bayesian confidence level, which corresponds to $\Delta \chi^2 = 4.6$ for a $\chi^2$ function depending on two parameters.

\subsection{Improvements of the EWPO}

\begin{table}
\begin{center}
\begin{tabular}{|l|cccccccc|}
\hline
 & $R_\ell$ & $R_{inv}$ & $R_b$ & $R_c$ & $M_W$ [MeV] & $s_{eff}^{2,\ell}$ & $\sigma_h^0$ [nb] & $\Gamma_\ell$ [MeV] \\
\hline
$\delta_{ILC}$ & 0.004  & 0.01  & 0.0002 & 0.0009 & 2.5  & 1.3 $\times 10^{-5}$ &  0.025  & 0.042  \\
$\delta_{FCC}$ & 0.001 & 0.002 & 0.00002 & 0.00009 & 0.5 & 1 $\times 10^{-6}$ &  0.0025 & 0.0042 \\
\hline
\end{tabular}
\end{center}
\caption{Projected precision of FCC-ee/TLEP and ILC for the high energy EWPOs. Estimates for FCC-ee/TLEP were taken from Ref.~\cite{Gomez-Ceballos:2013zzn}, and for the ILC from Ref.~\cite{Baak:2013fwa}. The improvement factor for $R_c$ was estimated to be identical to the one for $R_b$. We use the estimated systematic errors, noting that the achievable statistical errors are much smaller.}
\label{tab:tlep}
\end{table}

In this subsection, we consider only the EWPO, which depend on $\eps_{\tau\tau}$ and on the sum $\eps_{ee}+\eps_{\mu \mu}=:\eps_+$. We analyze the $\chi^2$ distribution under the assumptions~(\ref{assumption1}) and (\ref{assumption2}). 

Improved measurements of the EWPO are possible at the International Linear Collider (ILC), which will presumably produce $10^9$ $Z$ bosons in the so called GigaZ mode and at FCC-ee/TLEP, a circular electron positron collider with three times the circumference of LEP. The latter collider option allows a considerably higher luminosity compared to the former, which in turn allows the production of up to $10^{12}$ $Z$ bosons and $10^8$ $W$ bosons, which is referred to as the TeraZ and OkuW mode, respectively.
The error estimates used in this subsection for both collider options are listed in Tab.~\ref{tab:tlep}. We consider the projected systematic uncertainties for the two colliders rather than the statistical ones which would be much smaller. We note that, concerning the analysis for the two colliders, the theory uncertainties are set to zero (which will be discussed separately). 

\begin{figure}
\begin{minipage}{0.49\textwidth}
\begin{center}
\includegraphics[scale=0.6]{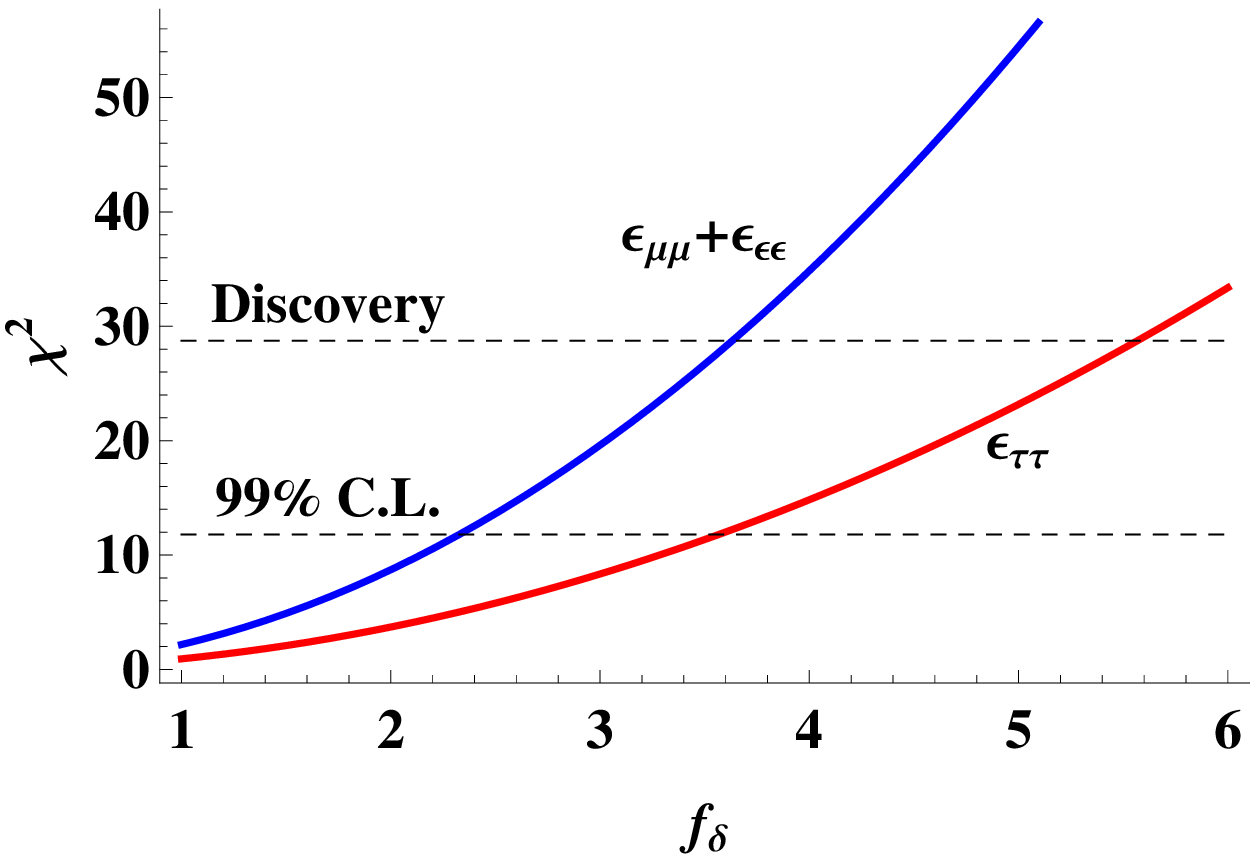}
\end{center}
\end{minipage}
\begin{minipage}{0.49\textwidth}
\begin{center}
\includegraphics[scale=0.5]{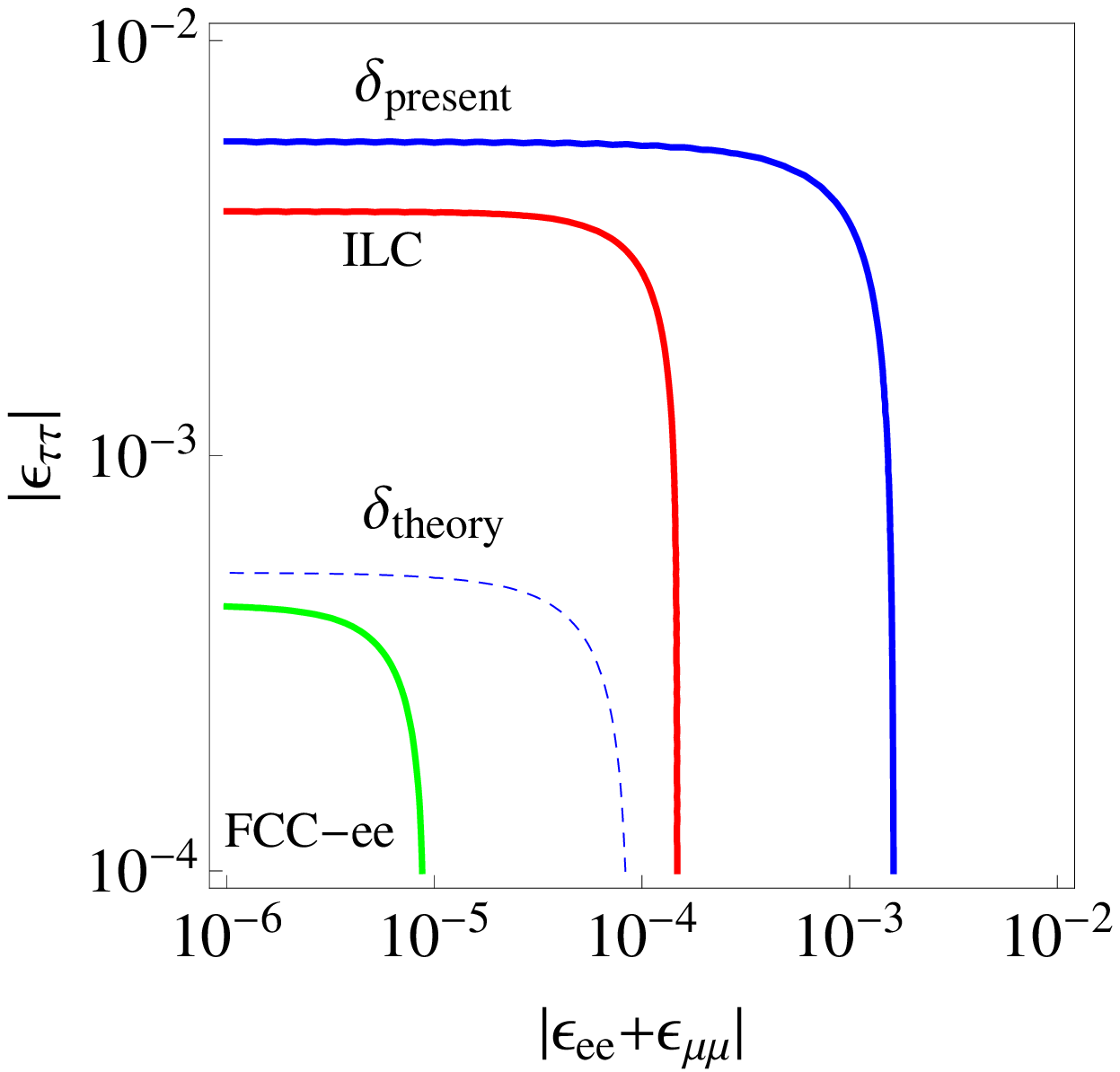}
\end{center}
\end{minipage}
\caption{{\it Left:} Improved precision needed at future EWPO measurements for a discovery of leptonic non-unitarity in $\eps_{ee}+\eps_{\mu\mu} =:e_+$  or $\eps_{\tau\tau}$. The x-axis denotes the improvement factor in experimental precision over the LEP measurements. {\it Right:} Exclusion limits on the $\eps_{\tau\tau}$ and $\eps_+$ at 90\% confidence level. The solid (dashed) blue line represents the current experimental (theoretical) uncertainty of the EWPO. The red and green lines denote the systematic ILC and FCC-ee/TLEP uncertainty, respectively. For details, see the text and Tab.~\ref{tab:tlep}.}
\label{fig:TLEP}
\end{figure}

The left panel of Fig.~\ref{fig:TLEP} shows the minimal $\chi^2$ of $\eps_+=0$ and $\eps_{\tau\tau}=0$, represented by the blue and red lines, respectively, as a function of $f_\delta$. The discovery of non-unitarity via each of the two independent parameters at $3\sigma$ and $5\sigma$ is denoted by the two dashed horizontal lines. The figure shows that the discovery of $\eps_+,\,\eps_{\tau\tau}$ requires an improvement factor $f_\delta \approx 3.5,\,5.0$, respectively. A comparatively modest experimental improvement of a factor $f_\delta \approx 2.6$ is already sufficient to exclude the SM, albeit without specifying explicitly which MUV parameter is non-zero.

The sensitivities to $\eps_+,\eps_{\tau\tau}$ of the ILC and FCC-ee/TLEP, respectively, are shown in the right panel of Fig.~\ref{fig:TLEP}, at the 90\% confidence level. The solid (dashed) blue line represents the current experimental (theoretical) uncertainties. The red and green line represent the 90\% exclusion sensitivity of the systematic FCC-ee/TLEP and ILC uncertainties, respectively. Fig.~\ref{fig:TLEP} shows that both the ILC and FCC-ee/TLEP have sufficient sensitivity to the MUV parameters to discover leptonic non-unitarity, provided the present best-fit values for the non-unitarity parameters are true. The figure clearly shows that in order to exploit the full FCC-ee/TLEP potential, some work on the theory uncertainties is necessary. The FCC-ee/TLEP sensitivity of $\sim 9 \times 10^{-6}$ can be translated via eq.~(\ref{eq:d=6coeffs}) into a mass for the ``sterile'' or ``right-handed'' neutrino of $\sim 60$ TeV, for Yukawa couplings of order one.

\subsection{Improved tests of lepton universality}
\begin{table}
\begin{center}
\begin{tabular}{|l|ccccc|}
\hline
Observable & $R^\ell$ & $R^\pi$ & $R^K$ & $Br(W\to\ell \nu)$ & $Br(W\to\ell \nu)$ \\
\hline
Precision & 0.001 & 0.001 & 0.004 & 0.0003 & 0.003 \\
Experiment & Tau factories & TRIUMF, PSI & NA62 & FCC-ee/TLEP & ILC \\
Reference & \cite{Biagini:2013fca} & \cite{AguilarArevalo:2010fv,Pocanic:2003pf} & \cite{Goudzovski:2010uk} & \cite{Gomez-Ceballos:2013zzn} & estimate \\
\hline
\end{tabular}
\end{center}
\caption{Estimated precision for future measurements of the lepton universality observables, from planned and running low energy experiments and FCC-ee/TLEP.}
\label{tab:LEimprovement}
\end{table}

Several experiments at the low energy or intensity frontier are being planned and commissioned, see Refs.~\cite{Erler:2014fqa,Kumar:2013qya} and references therein. In particular, planned and running low energy experiments will measure the lepton universality observables much more precisely. Furthermore, future lepton colliders such as ILC and FCC-ee/TLEP can strongly improve the universality measurements from W decays.

\begin{figure}
\begin{minipage}{0.49\textwidth}
\begin{center}
\includegraphics[width=0.95\textwidth]{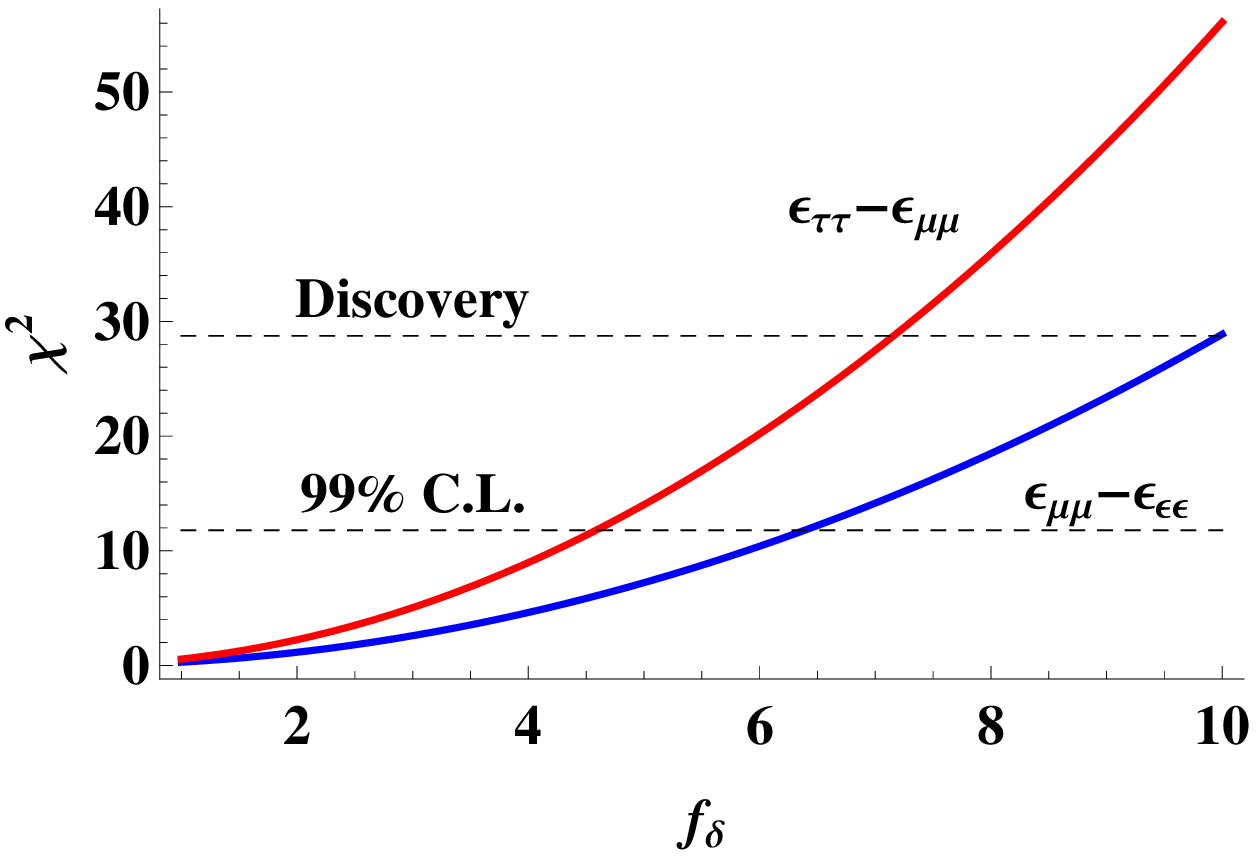}
\end{center}
\end{minipage}
\begin{minipage}{0.49\textwidth}
\begin{center}
\includegraphics[width=0.75\textwidth]{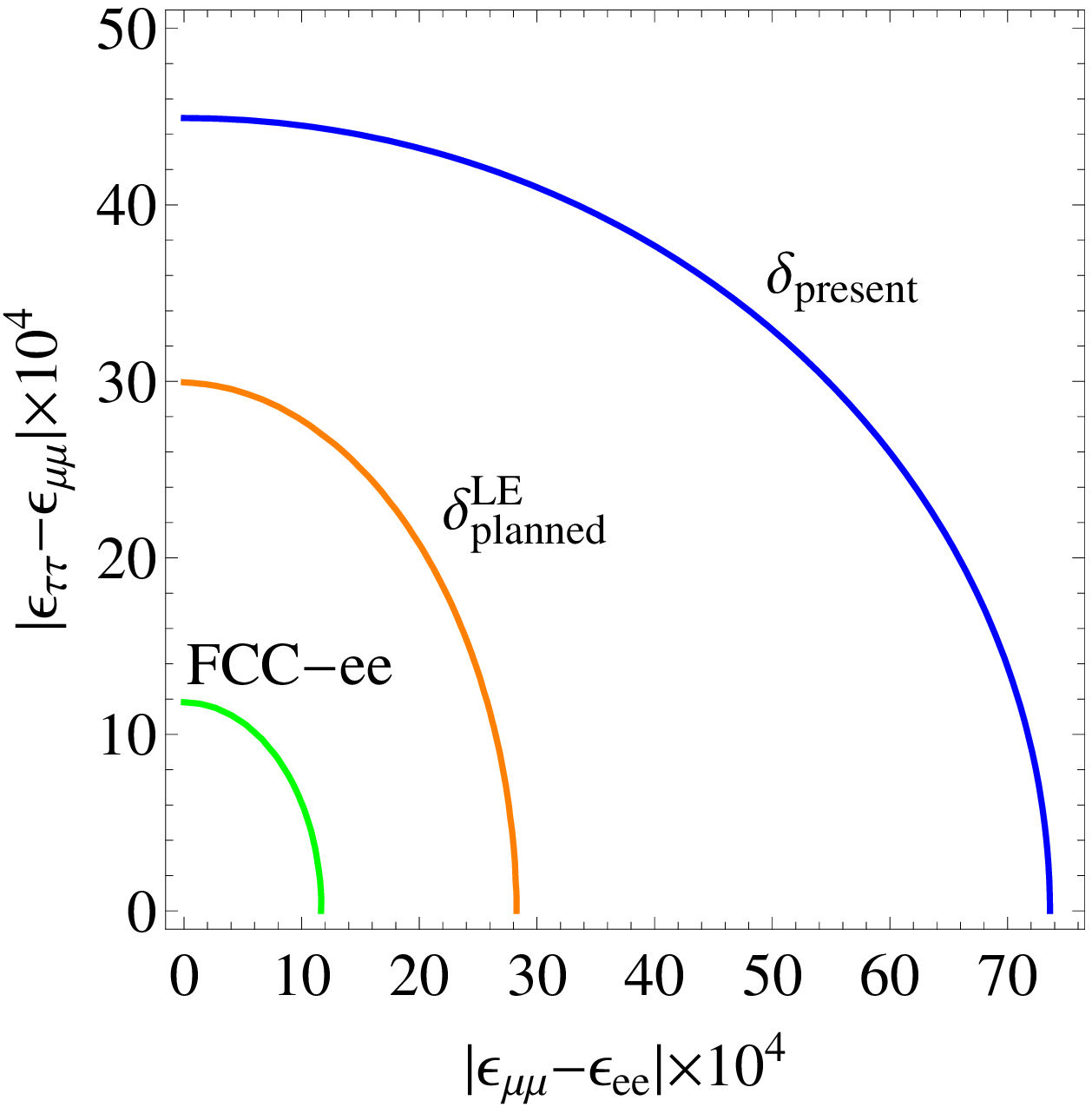}
\end{center}
\end{minipage}
\caption{{\it Left:} Improved precision needed at future universality test experiments for a discovery of leptonic non-unitarity in $\eps_{\tau\tau} -\eps_{\mu\mu}$ (red line) or $\eps_{\mu\mu}-\eps_{ee}$ (blue line). The horizontal dashed lines represent the statistical significance at 3$\sigma$ and 5$\sigma$, respectively. The x-axis denotes the improvement factor in experimental precision over the LEP measurements. {\it Right:} Exclusion limits at 90\% confidence level. The blue line denotes the present experimental exclusion limits, the orange line represents the estimated improvements in the low energy sector, see text and Refs.~\cite{Mammei:2012ph,AguilarArevalo:2010fv,Pocanic:2003pf,Goudzovski:2010uk,Biagini:2013fca}. The green line includes the estimated improvement of $W\to \ell_\alpha$ decays at the FCC-ee/TLEP. Sensitivities at ILC are very similar to the ones at low energy experiments (orange line).}
\label{fig:LEprecision}
\end{figure}

We present the estimated improvements for the universality observables in Tab.~\ref{tab:LEimprovement}. The precision of the ILC measurement for leptonic $W$ decays was estimated to be a factor 10 worse compared to FCC-ee/TLEP, due to the factor 100 smaller luminosity at this energy \cite{Gomez-Ceballos:2013zzn}. The MUV prediction for the universality observables depends on the differences between the diagonal epsilon parameters, such that the $\chi^2$ distribution depends only on two parameters. We choose the following differences as parameters:
\be
\Delta_{\tau\mu} := \eps_{\tau\tau} - \eps_{\mu\mu}\, \qquad \text{and}\qquad \Delta_{\mu e} := \eps_{\mu\mu} - \eps_{ee}\,.
\ee
The left panel of Fig.~\ref{fig:LEprecision} shows the growth of the minimal $\chi^2$ for $\Delta_{\tau\mu}=0$ and $\Delta_{\mu e}=0$, as a function of $f_\delta$. The two horizontal dashed lines denote the discovery of non-unitarity in leptonic mixing at the $3\sigma$ and $5\sigma$ level, respectively. We find that, under the assumption~(\ref{assumption1}), a discovery of non-unitarity in the parameters $\Delta_{\mu e},\,\Delta_{\tau \mu}$ is possible with an improved precision factor of $f_\delta \approx 10,\,7$, respectively. We note that the parameter $\Delta_{\tau \mu}$ requires a smaller improvement for a discovery compared to $\Delta_{\mu e}$ due to the best-fit value of the former being larger than that of the latter. For $f_\delta\approx 6$, the absence of non-unitarity can be excluded at the $5\sigma$ level without specifying which MUV parameter is non-zero. 

With the assumption~(\ref{assumption2}), the right panel of Fig.~\ref{fig:LEprecision} shows the exclusion sensitivity contour for the two MUV parameters at 90 \% CL. The blue line represents the current experimental precision while the improvements by low energy experiments are give by the orange line. The combination of the low energy improvements and leptonic $W$ decays from FCC-ee/TLEP, are represented by the green line. With the estimated ILC precision shown in Tab.~\ref{tab:LEimprovement}, the sensitivity to the MUV parameters is very similar to the one of the planned low energy experiments.

\subsection{Searches for charged lepton flavour violation}

The present experimental constraints on charged lepton flavour violation, discussed in section \ref{subsubsec:raredecays}, impose significant constraints on the off-diagonal non-unitarity parameters. While they provide the dominant constraint for $\eps_{e\mu}$, it is interesting to note that for the other two off-diagonal elements $\eps_{e\tau}$ and $\eps_{\mu\tau}$, the constraints from the triangle inequality given in eq.~(\ref{prior2}) comparable to the ones from experiment, as already discussed in section 4.2.

In the near future, rare tau decays could be improved up to $10^{-9}$ at SuperKEKB \cite{Akeroyd:2004mj}, which would probe the according off-diagonal $\eps$ parameters at the 0.1\% level. 

Furthermore, the experiments Mu3e at PSI \cite{Blondel:2013ia} and the proposed MUSIC project in Osaka \cite{Ogitsu:2011rg,Yamamoto:2011zb} are estimated to yield a sensitivity of $10^{-16}$ on the $\mu \to 3e$ branching ratio. The proposals for the Mu2e experiment at Fermilab \cite{Abrams:2012er} and the COMET experiment at J-PARC \cite{Kuno:2013mha} both estimate a sensitivity of $10^{-16}$ on the atomic $\mu \to e$ conversion rate. The PRISM/PRIME project \cite{Barlow:2011zza}, and a recent proposal \cite{Knoepfel:2013ouy} for an upgraded Mu2e experiment suggest, that a further improvement of this sensitivity up to $2 \times 10^{-18}$ is possible. This remarkable sensitivity would test $\eps_{\mu e}$ at the level of $3.6\times10^{-7}$ and $4.7\times10^{-7}$ for Titanium and Aluminium, respectively. With eq.~(\ref{eq:d=6coeffs}) this can be translated into the mass of the fermionic singlets of $\sim 0.3$ PeV, for neutrino Yukawa couplings of order one. We summarize the estimated sensitivities to the MUV parameters for these planned and proposed experiments in Tab.~\ref{tab-offdiag2}.

\begin{table}
\begin{center}
\begin{tabular}{|c|c|c|c|}
\hline
Process & MUV Prediction & Sensitivity (90 \% C.L.) & Bounds \\
\hline\hline
$Br_{\tau e}$ & $4.3\times 10^{-4}|\eps_{\tau e}|^2$  & $10^{-9}$ \cite{Akeroyd:2004mj} & $\eps_{\tau e} <1.5 \times 10^{-3}$ \\
$Br_{\tau \mu}$ & $4.1\times 10^{-4}|\eps_{\tau \mu}|^2$  & $10^{-9}$ \cite{Akeroyd:2004mj} & $\eps_{\tau \mu} <1.6 \times 10^{-3}$ \\
$Br_{\mu  eee}$ & $1.8 \times 10^{-5} |\eps_{\mu e}|^2$ & $10^{-16}$ \cite{Blondel:2013ia} & $\eps_{\mu e} < 2.4 \times 10^{-6}$\\
$R_{\mu e}^{Ti}$ & $1.5\times 10^{-5} |\eps_{\mu e}|^2$ & $2\times10^{-18}$ \cite{Knoepfel:2013ouy} & $\eps_{\mu e} < 3.6 \times 10^{-7}$\\
\hline
\end{tabular}
\end{center}
\caption{Projected future sensitivities for rare leptonic decays and muon-to-electron conversion in nuclei, cf. section \ref{subsubsec:raredecays}.}
\label{tab-offdiag2}
\end{table}

\subsection{Improved low energy measurements of $s_W^2$}
New experiments are planned at the intensity frontier, intended to measure the weak mixing angle at low energies, i.e.\ off the $Z$ peak.  
Already the final estimated precision of the total QWeak dataset, of which $\sim 4\%$ of the total data were discussed in section~\ref{weakchargeproton} and included in the present results, is expected to yield a relative precision of $\delta s_W^2 \sim 0.3\%$ \cite{Nuruzzaman:2013bwa}. 

Very encouraging in terms of accuracy is the P2 project in Mainz, which aims to measure the weak mixing angle via the parity violating asymmetry from elastical electron-proton scattering at low energy. A relative uncertainty for $s_W^2$ of $\sim 0.15\%$ is expected \cite{Becker:2013fya} around 2023. 
Yet a higher precision is estimated for the planned MOLLER experiment at Jefferson Laboratory, supposed to achieve $\delta s_W^2 \sim 0.125\%$, which would be five times more precise than SLAC-E158 \cite{Mammei:2012ph}.

Those experiments can furthermore help to resolve the discrepancy between the $Z$ pole observables $s_{W,\mathrm{eff}}^{\ell,\mathrm{lep}}$ and $s_{W,\mathrm{eff}}^{\ell,\mathrm{had}}$. A combination will yield a less ambiguous value of the weak mixing angle, with a higher precision.

\section{Summary and conclusions}\label{sec:summary}

We have performed a global fit to confront leptonic non-unitarity within the Minimal Unitarity Violation (MUV) scheme with the currently available experimental data. The results are given in section \ref{sec:constraints}, where we present the best-fit parameters as well as the allowed $1\sigma$ and $90\%$ Bayesian confidence level (CL) regions. 

We find that the data prefers flavour-conserving non-unitarity at 90\% CL for the parameter $\eps_{ee}$ and just below 90\% CL for the parameter $\eps_{\tau\tau}$. The moduli of the parameters $\eps_{\mu\mu}$ and $\eps_{\tau\tau}$ are constrained (at 90\% CL) to be $<0.0004$ and $<0.0053$, respectively. The flavour-violating parameter $\eps_{e\mu}$ is strongly constrained by the experimental results on the rare $\mu \to e \gamma$ decay. The experimental constraints for the other two flavour-violating parameters $\eps_{e \tau}$ and $\eps_{\mu \tau}$ are comparable to those from the triangle inequality in eq.~(\ref{prior2}), which is implied for MUV generated by SM extensions with fermion singlets.    

We have discussed the discovery potential and the exclusion sensitivity of various envisioned future experiments. 
Future lepton colliders such was the ILC and FCC-ee/TLEP would provide improved precision for the EWPO, which should be sufficient to turn the present ``hints'' for leptonic non-unitarity into a discovery, when the present best-fit non-unitarity parameters are assumed to be true. Under the assumption of unitarity, FCC-ee/TLEP could set bounds on the flavour conserving non-unitarity parameters, explicitly $\eps_{ee}+\eps_{\mu \mu} \leq 9 \times 10^{-6}$ and $\eps_{\tau\tau} \leq 4 \times 10^{-4}$. 
For leptonic non-unitarity generated by heavy sterile neutrinos with Yukawa couplings ${\cal O}(1)$, the strong constraint on $\eps_{ee}+\eps_{\mu \mu}$ would allow to test non-unitarity for sterile neutrino masses up to $\sim 60$ TeV.

Complementary information can be provided by future improved tests of lepton universality, either from planned low energy experiments or from $W$-decays at future lepton colliders. The experiments considered here are sensitive to the differences $\eps_{\alpha\alpha}-\eps_{\beta\beta}$ of the flavour-conserving non-unitarity parameters. Compared to the EWPO measurements, a larger improvement is necessary for a discovery (at the 5$\sigma$ level). Largest improvement could again be achieved at FCC-ee/TLEP. The sensitivity to the flavour changing non-unitarity parameters $|\eps_{\alpha\beta}|$ will be strongly improved by the planned experiments searching for rare tau and muon decays and by experiments on $\mu$-$e$ conversion in nuclei. In particular, $|\eps_{\mu e}|$ can be probed up to ${\cal O}(10^{-7})$, allowing to test non-unitarity from sterile neutrinos with masses up to $\sim 0.3$ PeV. We also discussed future low energy measurements of the weak mixing angle, which could help to resolve the present discrepancy between the leptonic and hadronic measurements from LEP.

Concluding our analysis, we state that searches for non-unitarity of the effective low energy leptonic mixing matrix constitute a useful test of whole classes of SM extensions, which account for the observed neutrino masses. Our global fit to the present data shows that non-zero $\eps_{ee}$ and $\eps_{\tau\tau}$ are preferred at about 90\% CL. Nevertheless, we rather interpret the present fit-results as bounds than as ``hints''. Future experiments will greatly improve the sensitivities such that either a clear discovery or strong bounds on the non-unitarity parameters can be achieved. To make full use of the estimated high precision of future EWPO measurements, it is important that also the theory uncertainties get improved accordingly.

\subsection*{Acknowledgements}
This work was supported by the Swiss National Science Foundation. We thank Alberto Luisiani and Tim Gershon for support with flavor data.

\section*{Appendix}
\begin{appendix}

\section{Asymmetry parameters}
\label{sec:asymmetry}
The fermion specific effective weak mixing angle is extracted from the experimentally measured forward-backward asymmetries $A_{FB}^f$  in $e^+ e^-$ collisions, which in turn can be related to the left-right asymmetries $A_{LR}^f$ by 
\be
A_{FB}^f = A_{LR}^e  A_{LR}^f \,,
\ee
where the $LR$--asymmetry parameters are given by
\be
A_{LR}^f = \frac{2 g_{V,f}  g_{A,f}}{g_{V,f}^2+g_{A,f} ^2}\,,
\ee
with the vector and axial couplings given by eq.~(\ref{eq:leptoncoupling}). 
Due to the different charge assignments between charged leptons, neutrinos, and up/down-quarks, the asymmetries acquire different values, and a different dependence on the matrix elements of $NN^\dagger$. We display the theory predictions of the MUV scheme and the SM for the asymmetry parameters, together with the experimental measurements in Tab.~\ref{tab:asymmetryparameters}. The MUV predictions are, as usual, approximated to first order in $\eps_{\alpha\beta}$.

\begin{table}
\begin{center}
$\begin{array}{|c|c|c|}
\hline
\text{Prediction in MUV} & \text{Prediction in the SM} & \text{Experiment} \\
\hline\hline
\left[A_\ell\right]_{\rm SM}\left(1-10.9(\eps_{ee}+\eps_{\mu\mu})\right) & 0.1475(10) & 0.1499(17) \\
\left[A_{FB}^\ell\right]_{\rm SM}\left(1 - 21.7 (\eps_{ee}+\eps_{\mu\mu})\right) & 0.0163(2) & 0.0171(10) \\
\left[A_b\right]_{\rm SM}\left(1 - 0.1 (\eps_{ee}+\eps_{\mu\mu})\right) & 0.9348(1) & 0.923(20) \\
\left[A_c\right]_{\rm SM}\left(1 - 0.85 (\eps_{ee}+\eps_{\mu\mu})\right) & 0.6680(4) & 0.670(27) \\
\left[A_{FB}^b\right]_{\rm SM}\left(1 - 9.2(\eps_{ee}+\eps_{\mu\mu})\right) & 0.1034(7) & 0.0992(16)\\
\left[A_{FB}^c\right]_{\rm SM}\left(1 - 10.0 (\eps_{ee}+\eps_{\mu\mu})\right) & 0.0739(5) & 0.0707(35)\\
\hline
\end{array}$
\end{center}
\caption{Experimental results, Standard Model predictions and modifications in MUV for the electroweak asymmetry observables.}
\label{tab:asymmetryparameters}
\end{table}

In the analysis presented in the main part of the paper, we have not included the individual asymmetry observables but rather the two experimental results for the effective weak mixing angle, extracted from leptonic and hadronic measurements as shown in Table~\ref{precisionobservables}. We have checked that using alternatively the individual asymmetries does not significantly change the results of our analysis.

\section{Correlations}\label{sec:correlations}
In our analysis the correlations between observables are included via the correlation matrix $R$, which has the correlations as off-diagonal elements and $1$ as diagonal entries. The following correlation coefficents are used:
\begin{center}
\begin{tabular}{cc|c|c}
\multicolumn{2}{c}{Correlated } & coefficient & Ref. \\
\hline
$R^W_{\tau \mu}$ & $R^W_{\mu e}$ & 0.44 & \cite{Schael:2013ita} \\
$\Gamma_{lept}$ & $R_{inv}$ &  0.17 & \cite{ALEPH:2005ab}  \\
$\sigma_{had}$ & $R_\ell$ & 0.183 & \cite{ALEPH:2005ab} \\
\end{tabular}
\end{center}
The correlation between the hadronic EWPO in percent is given by \cite{ALEPH:2005ab}:
\be
\begin{array}{c|rrrrr}
 & A_c & A_{FB}^b & A_{FB}^c & R_b & R_c \\
\hline
A_b &  11 &  6 &  -2 &  -8 &  4\\ 
A_c &   &  1 &  4 &  4 &  -6\\ 
A_{FB}^b &   &   &  15 &  -10 &  4\\ 
A_{FB}^c  &   &   &   &  7 &  -6\\ 
R_b  &   &   &   &   &  -18\\ 
\end{array}
\nonumber
\ee
The correlation between the CKM observables in percent reads \cite{Antonelli:2010yf}:
\be
\begin{array}{c|cccc}
 & K_L \to \mu & K_S \to e & K^\pm \to e & K^\pm \to \mu \\
\hline
K_L \to e &  55 &  10 &  3 &  0\\ 
K_L \to \mu &   &  6 &  0 &  4\\ 
K_S \to e &   &   &  1 &  0\\  
K^\pm \to e &   &   &   &  73\\  
\end{array}
\nonumber
\ee
The correlation between the lepton universality observables in percent is \cite{Amhis:2012bh,HFAGwebsite}:
\be
\begin{array}{c|cccc}
  & R_{\tau \mu}^\ell & R_{\tau e}^\ell & R_{\mu e}^\ell & R_{\tau \mu}^\pi \\
\hline
R_{\tau e}^\ell & 77 & & &  \\
R_{\mu e}^\ell & -35 & 34 & &  \\
R_{\tau \mu}^\pi & 49 & 50 & 2 & \\
R_{\tau \mu}^K & 23 & 21 & -2 & 14 
\end{array}
\nonumber
\ee

\end{appendix}

\bibliographystyle{unsrt}

%\bibliography{library}

\end{document}